# Hot electrons and electromagnetic effects in the broadband Au, Ag, and Ag-Au nanocrystals: The UV, visible, and NIR plasmons

Alina Muravitskaya,[1,2,‡] Artur Movsesyan,[1,2,‡,*] Oscar Ávalos-Ovando,[2] Verónica A. Bahamondes Lorca,[3,4] Miguel A. Correa-Duarte,[5] Lucas V. Besteiro,[5] Tim Liedl,[6] Peng Yu,[1] Zhiming Wang,[1] Gil Markovich,[7] and Alexander O. Govorov[2,*]

[1] Institute of Fundamental and Frontier Sciences, University of Electronic Science and Technology of China, Chengdu 610054, China

[2] Department of Physics and Astronomy and Nanoscale and Quantum Phenomena Institute, Ohio University, Athens 45701, OH, USA

[3] Edison Biotechnology Institute, Ohio University, Athens 45701, OH, USA

[4] Departamento de Tecnología médica, Facultad de Medicina, Universidad de Chile, Santiago, Chile

[5] CINBIO Universidade de Vigo Vigo 36310, Spain

[6] Faculty of Physics and Center for NanoScience (CeNS), Ludwig Maximilians University, 80539 Munich, Germany

[7] School of Chemistry, Raymond and Beverly Sackler Faculty of Exact Sciences, Tel Aviv University, Tel Aviv, 6997801, Israel

* Artur Movsesyan, e-mail: movsesyan@gmail.com ; Alexander O. Govorov, e-mail: govorov@ohio.edu

‡These authors contributed equally

ABSTRACT. Energetic and optical properties of plasmonic nanocrystals strongly depend on their sizes, shapes, and composition. Whereas using plasmonic nanoparticles in biotesting has become a routine, applications of plasmonics in energy are still early in development. Here, we investigate hot electron (HE) generation and related electromagnetic effects in both mono- and bi-metallic nanorods (NRs) and focus on one promising type of bi-metallic nanocrystals – core-shell Au-Ag nanorods. The spectra of the NRs are broadband, highly tunable with their geometry, and have few plasmon resonances. In this work, we provide a new quantum formalism describing the HE generation in bi-metallic nanostructures. Interestingly, we observe that the HE generation rate at the UV plasmon resonance of Au-Ag NRs appears to be very high. These HEs are highly energetic and suitable for carbon-fuels reactions. Simultaneously, the HE generation at the longitudinal plasmon (L-plasmon) peaks, which can be tuned from the yellow to near-IR, depends on the near-field and electromagnetic Mie effects, limiting the HE efficiencies for the long and large NRs. These properties of the L-plasmon relate to all kinds of NRs (Au, Ag, and Au-Ag). We also consider the generation of the interband d-holes in Au and Ag, since the involvement of the d-band is crucial for the energetic properties of UV plasmons. The proposed formalism is an important development for the description of bi-metallic (or tri-metallic, or more complex) nanostructures, and it paves the way to the efficient application of the plasmonic HEs and hot holes in sensing, nanotechnology, photocatalysis, and electrophotochemistry.

INTRODUCTION

Nowadays, plasmonic nanocrystals are being routinely used in lateral flow assay biotesting strips designed for the naked-eye detection of various biomarkers such as Covid proteins, signaling and differentiating molecules for heart attack (e.g., troponin), stroke or cerebral edema, etc.[1,2] Simultaneously, applications of plasmonic nanostructures for energy harvesting and photosynthesis of fuels remain under investigation, although being heavily studied.[3–6] The effect of generating energic hot electrons (HEs) indeed looks very promising when a solar light energy quantum creates an electron-hole pair in the Fermi sea.[7,8] HEs typically have energies much higher than the thermal energy of the material, often many times higher. Another thing about HEs is that they exist in a non-equilibrium distribution, i.e., strongly differing from the standard Fermi-Dirac distribution of electrons at thermal equilibrium. The advantages of plasmonic nanocrystals are their high absorption cross-sections and strongly amplified fields, whereas a fundamental obvious drawback lies in the ultra-short lifetimes of excited electrons.[7–9] One efficient route leading to the HE generation, which takes full advantage of electromagnetic enhancement in a metal nanocrystal, is the surface-scattering mechanism, also regarded as Kreibig's mechanism.[10] In this mechanism, the surface, being a strong lattice defect, allows breaking the linear-momentum conservation and an excitation of high-energy electron (Figure 1a). Such energetic HEs can be injected to a semiconductor (Schottky photodiode) or a surface molecule (photocatalysis). Along with HEs and hot holes (HHs) generated in the conduction (sp) band of a noble metal, optical excitation leads to the excitation of low-energy Drude electrons, generating plasmonic currents, and high-energy excitations of d-holes.[11–13] Our formalism, which combines the Kreibig model and our microscopic quantum theories,[8,11,14] allows us to compute the above contributions (Drude, HEs, and d-holes) separately. For this, we apply our nonlinear theory of HE generation, which utilizes the COMSOL and a nonlinear iterative approach.[11]

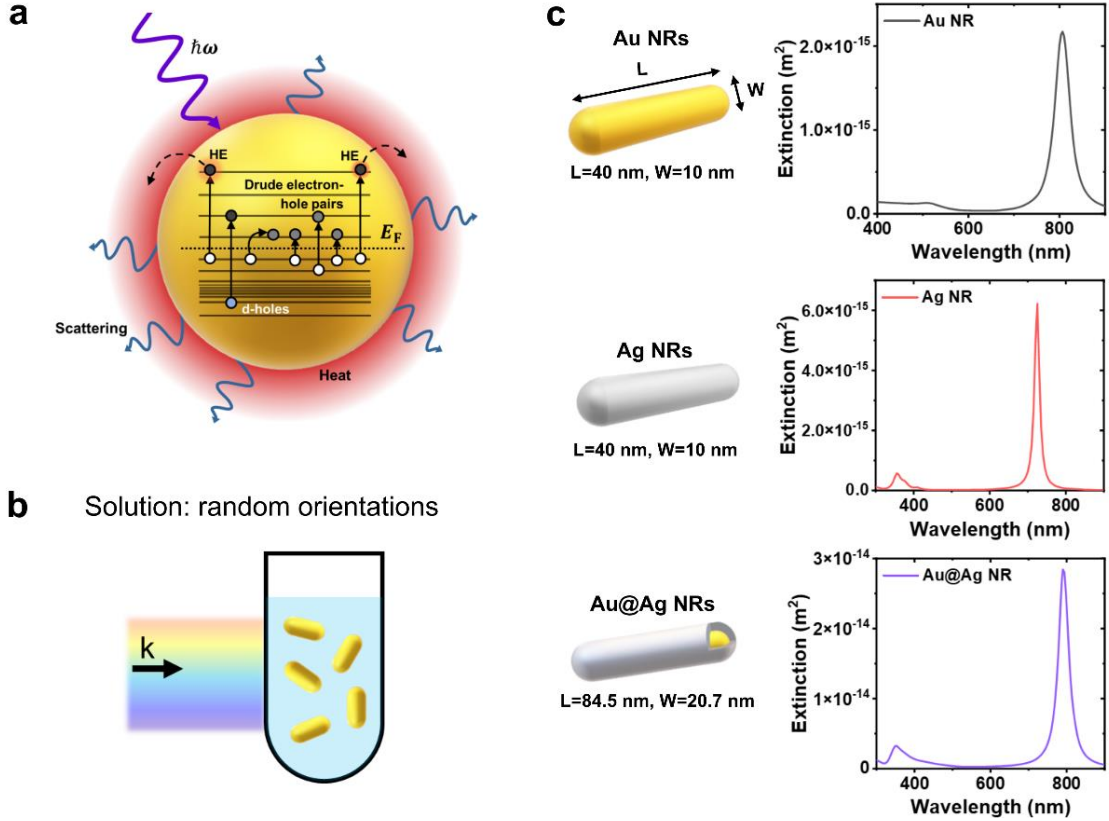

**Figure 1.** (a) Absorption, hot-electron, and photoheating processes in an optically driven plasmonic nanocrystal. (b) Model of randomly oriented NRs, i.e., a solution setting. (c) Extinction cross-sections of gold (W=10 nm, L=40 nm), silver (W=10 nm, L=40 nm) and Au@Ag NRs (W=20.7 nm, L=84.5 nm) with comparable aspect ratio.

Using our quantum formalism, we investigate here the HE generation and electromagnetic Mie effects in bi-metallic nanorods (NRs) and reveal a number of important features:

(1) We focus on the Ag-coated Au NRs since such a system has been recently mastered by several labs and exhibits broadband and highly tunable spectra.[15–18] Their spectra possess sharp plasmon resonances located in the spectral interval extending from UV to mid-IR.[17,18] The HE generation in such nanostructures is efficient[17] and can be used in a variety of applications ranging from energy harvesting to biosensing.

(2) In the Au@Ag NRs, we observe a rather unique feature - a strong plasmon resonance in the UV at 350 nm. This resonance is a slab-like oscillation in the Ag shell (T-plasmon).

(3) Why is the 350 nm resonance interesting? The energetic hot carries created in this resonance are of ~ 3.5 eV, such UV electrons can be used for carbon-based fuel catalysis.

(4) The rate of generation of UV HEs is high, comparable with the one at the major longitudinal resonance (L-plasmon) in the red.

(5) In the hybrid NRs, the HE generation occurs at both interfaces (Ag-water and Au-Ag).

(6) The electromagnetic Mie effects play an important role in the HE generation for the L-plasmon peaks, limiting the HE efficiencies. Here, we develop a theory for this Mie-like regime too.

(7) Finally, we pay attention to the generation of the interband d-holes in Au and Ag. For the UV plasmons, the involvement of the d-band is crucial.

We highlight that Au@Ag NRs have been synthesized and studied at this moment by many experimental groups[16–21] motivated by the unique possibility to utilize the plasmonic properties of silver in stable NRs.[17,18] We note that a wet chemistry fabrication of purely Ag NRs is more challenging and often does not provide nanocrystals stable over a long time.[22] Overall, Ag-incorporated nanocrystals with non-spherical shapes,[22–25] due to their strong plasmon resonances and spectral tunability, are an active ongoing field of research, where we should expect more progress towards stable Ag nanostructures. Here, we would like to present a large volume of photophysical data on the stable Au@Ag core-shell NRs, providing an understanding of plasmonic resonances (also in the UV!) and the HE generation processes in this new, exciting class of nanomaterials. For example, our study of Ref. [18] revealed the ultra-strong generation of HEs in the UV at about 350 nm, and an explanation for this observation has been missing so far. This paper explains it well, involving the slab-like Ag-based plasmon resonance. In addition, we provide a new formalism with $2\text{-}\gamma_s$ for a bi-metal nanostructure which was never done before. It is an important development for the description of bi-metallic (or tri-metallic, or even more complex) nanostructures.

**Nonlinear hot-electron formalism.**

**Optical properties.** The external field of a monochromatic electromagnetic wave exciting a metal nanocrystal (NC) can be written in the standard way:

$$\mathbf{E}_{ext} = \text{Re}\left[\mathbf{E}_0 \cdot e^{-i\omega t}\right] = \mathbf{E}_0 \cos[\omega t], \tag{1}$$

here $\mathbf{E}_0$ is a real amplitude. For this choice of the complex exponent in $\mathbf{E}_{ext}$, we should assume that $\text{Im}[\varepsilon_{metal}] > 0$; we should note that COMSOL Multiphysics is using the opposite convention: $\text{Re}\left[\mathbf{E}_0 \cdot e^{+i\omega t}\right]$. Correspondingly, the monochromatic local electric field inside and outside of a NC

$$\mathbf{E}(\mathbf{r}) = \text{Re}[\mathbf{E}_\omega(\mathbf{r}) \cdot e^{-i\omega t}] \tag{2}$$

where $\mathbf{E}_\omega(\mathbf{r})$ is the complex amplitude of the electric field at position $\mathbf{r}$ and for frequency $\omega$. The optical power absorbed inside a metal NC:[26]

$$P_{abs} = \left\langle \int_{NP} dV \, \mathbf{j} \cdot \mathbf{E} \right\rangle_{time} = \text{Im}[\varepsilon_{metal}] \cdot \varepsilon_0 \frac{\omega}{2} \int_{NC} dV \, \mathbf{E}_\omega \cdot \mathbf{E}_\omega^* \qquad (3)$$

where $\mathbf{j}$ is the density of current, and $\varepsilon_{metal}$ is the dielectric constant of the NC, described in detail below. The integral is taken over the volume of an absorbing object. For the absorption cross-section of a NC, we get

$$\sigma_{abs}(\omega) = \frac{P_{abs}}{I_0} \qquad (4)$$

$$I_0 = \frac{\varepsilon_0 c_0 \sqrt{\varepsilon_{env}}}{2} \cdot E_0^2 \qquad (5)$$

where $I_0$ is the energy flux of the incident electromagnetic wave, $c_0$ is the light speed in vacuum, $\varepsilon_{env}$ is the dielectric constant of the surrounding environment. In our numerical computation, we use water, $\varepsilon_{env} = 1.8$, and assume a Lambert-Beer experiment, i.e., our NCs are dispersed at random orientations (Figure 1b). The incident flux intensity is taken to be 400 W/cm² for all computations done below. Following the textbooks, for a NC, the total extinction power, $P_{tot}$, contains two terms:

$$P_{tot} = P_{scat} + P_{abs} \qquad (6)$$

Here $P_{scat}$ and $P_{abs}$ are the scattering and absorption rates, respectively. Then, the extinction cross-section reads:

$$\sigma_{ext}(\omega) = \frac{P_{ext}}{I_0} \qquad (7)$$

$$\sigma_{ext}(\omega) = \sigma_{scat}(\omega) + \sigma_{abs}(\omega)$$

where $\sigma_{scat}(\omega)$ and $\sigma_{abs}(\omega)$ are the scattering and absorption cross-sections, respectively. In our study, we look at both small and large sizes. For small NCs, $\sigma_{scat} \ll \sigma_{abs}$, whereas for large sizes, we enter the Mie regime, i.e., $\sigma_{scat} \sim \sigma_{abs}$. Any optical property in the solution setting should be averaged over 12 configurations,

$$\sigma_{av} = \frac{\sigma_{(k_x,E_y)} + \sigma_{(k_x,E_z)} + \sigma_{(k_y,E_x)} + \sigma_{(k_y,E_z)} + \sigma_{(k_z,E_x)} + \sigma_{(k_z,E_y)} + \sigma_{(-k_x,E_y)} + ...}{12} \qquad (8)$$

where $\mathbf{k}_\alpha$ are the wavevectors oriented along the three orthogonal axes, and $\mathbf{E}_\alpha$ are the corresponding electric-field polarizations of the incident light. Figure 1c shows the computed spectra of Au, Ag and Au@Ag NRs randomly dispersed in a solution.

**Hot carrier mechanisms.** Since we are interested in the HE generation, we should look at the physical mechanism contributing to the absorbed power, $P_{abs}$. Considering the main microscopic mechanisms of dissipation in a plasmonic NC (Figure 2c,d), we have:[11,27,28]

$$P_{abs} = P_{Drude} + P_{interband} + P_s \qquad (9)$$

where $P_{Drude}$, $P_s$, and $P_{interband}$ are the Drude, surface-scattering (Kreibig), and interband (d-band) contributions, correspondingly. Each term in Equation (9) describes a specific pathway of energy dissipation in a metal NC (Figure 1a). Ref. [11] describes those mechanisms in detail. Next, we express the dielectric constant a metal (Au or Ag) in the following way:

$$\varepsilon_{metal}(\omega) = 1 + \Delta\varepsilon_{interband}(\omega) - \frac{\omega_p^2}{\omega(\omega + i[\gamma_D + \gamma_s])},$$

$$\Delta\varepsilon_{interband}(\omega) = \varepsilon_{metal,bulk}(\omega) - 1 + \frac{\omega_p^2}{\omega(\omega + i\gamma_D)},$$

(10)

where $\varepsilon_{metal,bulk}(\omega)$ is the experimental dielectric constant for the bulk material (Au or Ag).[29] Furthermore, where $\gamma_D$ is the Drude broadening parameter, and $\omega_p$ is the plasmon frequency; $\omega_p = \sqrt{\frac{e^2 n_0}{\varepsilon_0 m}}$, in which $n_0$, $m$ and $\varepsilon_0$ are the electron density, the electron's mass, and the dielectric permittivity of free space, respectively. $\gamma_s$ is the Kriebig's parameter,[10] i.e., the effective plasmonic broadening due to the electron scattering mediated by the surface; this parameter is of a quantum nature.

For a bi-metallic object, the equations to compute the contributions are given by the integrals:

$$P_{interband,\beta} = \sum_{\beta=Au,Ag} \text{Im}\,\Delta\varepsilon_{interband,\beta}(\omega) \cdot \varepsilon_0 \frac{\omega}{2} \int_{V_\beta} dV\, \mathbf{E}_{\omega,\beta} \cdot \mathbf{E}_{\omega,\beta}^*$$

$$P_{Drude,\beta} = \sum_{\beta=Au,Ag} \frac{\omega_{p,\beta}^2}{\omega^3}(\gamma_{D,\beta}) \cdot \varepsilon_0 \frac{\omega}{2} \int_{V_\beta} dV\, \mathbf{E}_{\omega,\beta} \cdot \mathbf{E}_{\omega,\beta}^* \quad (11)$$

$$P_{s,\beta} = \sum_{\beta=Au,Ag} \frac{\omega_{p,\beta}^2}{\omega^3}(\gamma_{s,\beta}) \cdot \varepsilon_0 \frac{\omega}{2} \int_{V_\beta} dV\, \mathbf{E}_{\omega,\beta} \cdot \mathbf{E}_{\omega,\beta}^*$$

The key equations (11) were derived under the assumption $\omega^2 \gg (\gamma_D + \gamma_s)^2$, which applies well to the usual plasmonic NCs.[11] The quantum dissipation parameter $\gamma_s$ is size- and shape-dependent. This parameter is found computationally from a nonlinear integrodifferential equations, self-consistently.[11] The method was developed in Ref. [11,27] and more details can be found in the Supporting Information.

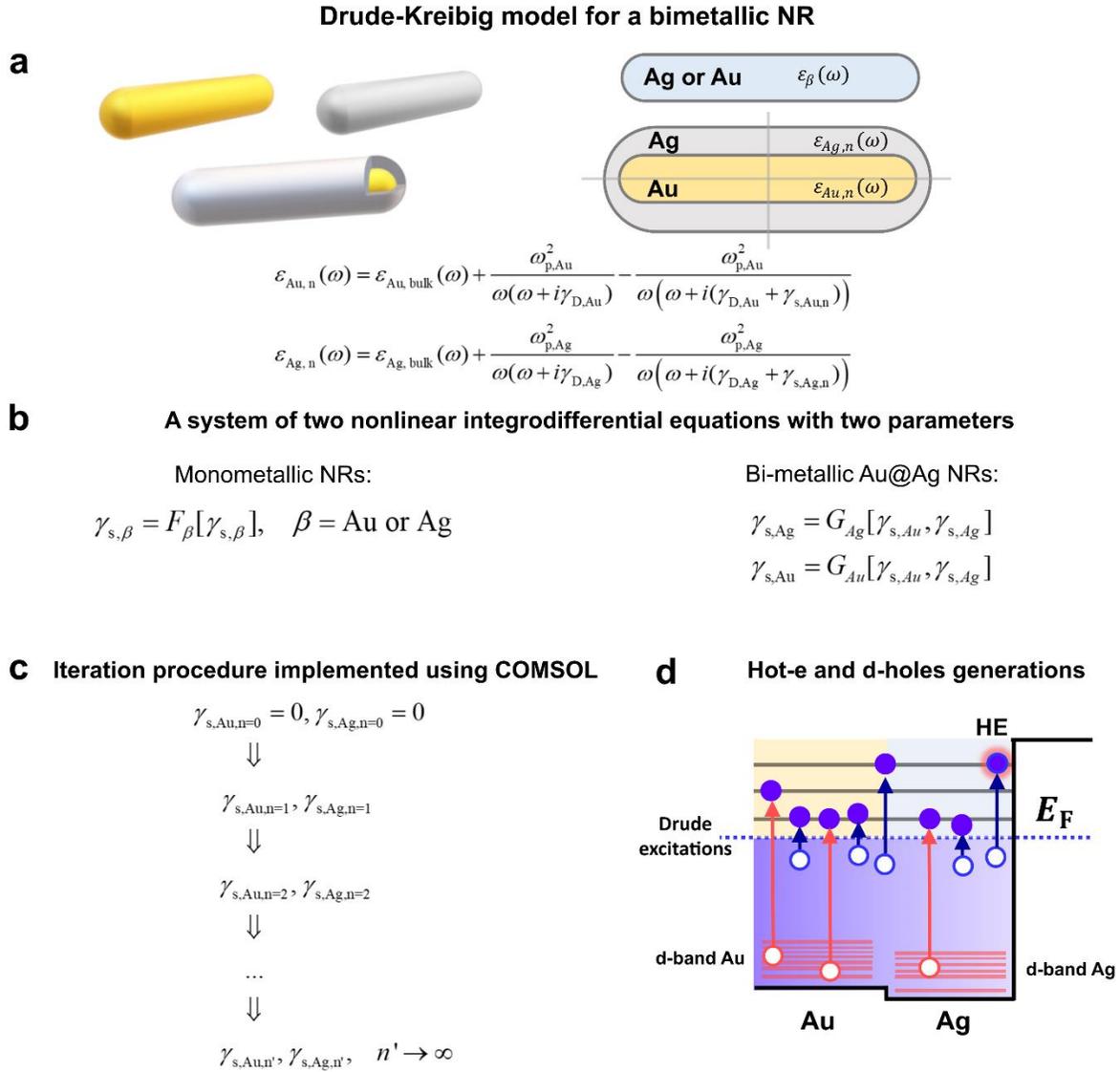

**Figure 2.** Drude-Kreibig model for a bimetallic nanocrystal and the iterative theory, developed by us in Refs.[11,12], solving a set of nonlinear integrodifferential equations for a NC with complex boundaries. (a-c) Equations for Drude-Kreibig model for a bimetallic NR. The Functionals, F and G, are given in the text. (d) HE generation in the sp-bands of Au and Ag in Au@Ag nanocrystals.

**Nonlinear integrodifferential formalism.** Fig. 2 illustrates the nonlinear formalism for the Kreibig parameters, $\gamma_{s,Au}$ and $\gamma_{s,Ag}$. The HE generation rate ($Rate_{HE}$) defines the total number of HEs excited inside a NC per second – the generation occurs at both interfaces (Au-Ag and Ag-water) as shown in Fig. 2d. The second parameter ($Rate_{high-E\ HE}$) is defined for the high-energy HEs, i.e., for the rate of excited electrons having over-barrier energies.[11,30] These electrons can propagate or tunnel to a semiconductor (if attached to a metal NC) or to a molecular state at the surface.

$$Rate_{HE} = \frac{P_{HE}}{\hbar\omega} = \sum_{Au-Ag, Ag-matrix} \frac{1}{4} \times \frac{2}{\pi^2} \times \frac{e^2 E_F^2}{\hbar} \frac{1}{(\hbar\omega)^3} \int_{S_{NP}} |E_{\omega,normal}(\theta,\varphi)|^2 \, ds$$

$$Rate_{high\text{-}E\,HE} = \frac{(\hbar\omega - \Delta E_{bar})}{(\hbar\omega)} \times \frac{P_{HE}}{\hbar\omega} \tag{12}$$

where $E_{\omega,\,normal}(\theta,\varphi)$ is the normal component of the electric field on the interfaces, inside the NC; $E_{F,Au} = 5.43$ eV and $E_{F,Ag} = 5.75$ eV are the Fermi energies. The integrals are taken over the surface since the HEs are being generated due to surface scattering; $\Delta E_{bar}$ is the injection barrier for the metal-matrix interface. In our computations, we employ $\Delta E_{bar} = 1$ eV, which is a typical number for the Au/Ag-TiO$_2$ Schottky-barriers. In bi-parametric nonlinear formalism, the surface-scattering HE generation is also expressed via the quantum dissipation parameters (Supporting Information):

$$Rate_{HE} = \frac{P_{HE}}{\hbar\omega} = \sum_{Au-Ag, Ag-matrix} \frac{1}{4} \times \frac{2}{\pi^2} \times \frac{e^2 E_F^2}{\hbar} \frac{1}{(\hbar\omega)^3} \int_{S_{NC}} |E_{\omega,normal}(\theta,\varphi)|^2 \, ds =$$

$$= \frac{1}{\hbar\omega} \sum_{\beta=Au,Ag} \frac{\omega_{p,\beta}^2}{\omega^3} (\gamma_{s,\beta}) \cdot \varepsilon_0 \frac{\omega}{2} \int_{V_\beta} dV \, \mathbf{E}_{\omega,\beta} \cdot \mathbf{E}^*_{\omega,\beta} \tag{13}$$

As one can see from Equation (12), $Rate_{high\text{-}E\,HE}$ describes the generation of electrons with high (over-barrier) energies, i.e., with $E > \Delta E_{bar}$. In a solution setting, the rates (Equation (12)) should be averaged over the twelve light configurations.

$$Rate_{HE} = \frac{1}{4} \times \frac{2}{\pi^2} \times \frac{e^2 E_F^2}{\hbar} \frac{1}{(\hbar\omega)^3} \int_{S_{NC}} |E_{\omega,normal}(\theta,\varphi)|^2 \, ds = \frac{P_{HE}}{\hbar\omega} = \frac{1}{\hbar\omega} \frac{\omega_p^2}{\omega^3} (\gamma_s) \cdot \varepsilon_0 \frac{\omega}{2} \int_{NC} dV \, \mathbf{E}_\omega \cdot \mathbf{E}^*_\omega$$

$$Rate_{high\text{-}E\,HE} = \frac{1}{4} \times \frac{2}{\pi^2} \times \frac{e^2 E_F^2}{\hbar} \frac{(\hbar\omega - \Delta E_{bar})}{(\hbar\omega)^4} \int_{S_{NC}} |E_{\omega,normal}(\theta,\varphi)|^2 \, ds, \tag{14}$$

The d-hole generation process, which is a bulk mechanism, is given as the local rate (in the units of 1/s*m$^3$):

$$r_{d\text{-holes, bulk},\beta}(\mathbf{r}) = \frac{d^2 N_{d\text{-holes}}}{dt dV} = \frac{1}{\hbar\omega} \text{Im}[\Delta\varepsilon_{interband,\beta}] \varepsilon_0 \frac{\omega}{2} \mathbf{E}_{\omega,\beta} \cdot \mathbf{E}^*_{\omega,\beta} \tag{15}$$

Correspondingly, the total d-hole rates are given by the volume integrals:

$$Rate_{d\text{-holes, bulk},\beta} = \sum_{\beta=Au,Ag} \varepsilon_0 \frac{1}{\hbar\omega} \text{Im}\,\Delta\varepsilon_{interband,\beta}(\omega) \frac{\omega}{2} \int_{V_\beta} dV \, \mathbf{E}_{\omega,\beta} \cdot \mathbf{E}^*_{\omega,\beta}$$

$$P_{d\text{-holes, bulk},\beta} = \sum_{\beta=Au,Ag} \text{Im}(\varepsilon_{interband,\beta}) \varepsilon_0 \frac{\omega}{2} \int_{V_\beta} dV \, \mathbf{E}_{\omega,\beta} \cdot \mathbf{E}^*_{\omega,\beta} \tag{16}$$

Finally, we look at the Drude-dissipation power:

$$P_{\text{Drude, bulk}}(\mathbf{r}) = \sum_{\beta=\text{Au,Ag}} \frac{\omega_{p,\beta}^2 \cdot \gamma_{D,\beta}}{\omega^3} \varepsilon_0 \frac{\omega}{2} \int_{V_\beta} dV \mathbf{E}_{\omega,\beta} \cdot \mathbf{E}_{\omega,\beta}^* \qquad (17)$$

Supporting Information contains all physical parameters used in the above equations and the detailed derivation for a bi-metallic NC, including the derivations for the nonlinear equations:

Monometallic NRs:

$$\gamma_{s,\beta} = \frac{3}{4} \times v_{F(\beta)} \frac{S_\beta}{V_\beta} \frac{F_{S,\beta,\text{ext}}(\gamma_{s,\beta,\text{n'-1}}, \gamma_{s,\beta,\text{n'-1}})}{F_V(\gamma_{s,\beta,\text{n'-1}}, \gamma_{s,\beta,\text{n'-1}})}$$

$\beta = $ Au or Ag

Bi-metallic Au@Ag NRs:

$$\gamma_{s,\text{Ag}} = \frac{3}{4} \times v_{F(\text{Ag})} \frac{S_{\text{Ag}}}{V_{\text{Ag}}} \frac{F_{S,\text{Ag,ext}}(\gamma_{s,\text{Au}}, \gamma_{s,\text{Ag}})}{F_V(\gamma_{s,\text{Au}}, \gamma_{s,\text{Ag}})} + \frac{|\Delta E_F|}{\overline{E}_F} \frac{1}{2} \left( \frac{3}{4} \times v_{F(\text{Au})} \frac{S_{\text{Au}}}{V_{\text{Au}}} \frac{F_{S,\text{Au-}\varepsilon}(\gamma_{s,\text{Au}}, \gamma_{s,\text{Ag}})}{F_{V,\text{Au}}(\gamma_{s,\text{Au}}, \gamma_{s,\text{Ag}})} + \frac{3}{4} \times v_{F(\text{Ag})} \frac{S_{\text{Ag}}}{V_{\text{Ag}}} \frac{F_{S,\text{Ag+}\varepsilon}(\gamma_{s,\text{Au}}, \gamma_{s,\text{Ag}})}{F_{V,\text{Ag}}(\gamma_{s,\text{Au}}, \gamma_{s,\text{Ag}})} \right)$$

$$\gamma_{s,\text{Au}} = \frac{|\Delta E_F|}{\overline{E}_F} \frac{1}{2} \left( \frac{3}{4} \times v_{F(\text{Au})} \frac{S_{\text{Au}}}{V_{\text{Au}}} \frac{F_{S,\text{Au-}\varepsilon}(\gamma_{s,\text{Au}}, \gamma_{s,\text{Ag}})}{F_{V,\text{Au}}(\gamma_{s,\text{Au}}, \gamma_{s,\text{Ag}})} + \frac{3}{4} \times v_{F(\text{Ag})} \frac{S_{\text{Ag}}}{V_{\text{Ag}}} \frac{F_{S,\text{Ag+}\varepsilon}(\gamma_{s,\text{Au}}, \gamma_{s,\text{Ag}})}{F_{V,\text{Ag}}(\gamma_{s,\text{Au}}, \gamma_{s,\text{Ag}})} \right)$$

Figure 2 illustrates the computation scheme with the F- and G-functionals given by the above equations.

To conclude this technical section, we note that our approach is COMSOL-based and directly involves the band structures of the constituting metals via their local dielectric constants $\varepsilon_{\text{metal,Ag}}(\omega)$ and $\varepsilon_{\text{metal,Au}}(\omega)$. Therefore, our theory allows us to compute colloidal NCs with arbitrarily complex shapes/compositions and lithographic metamaterials with arbitrary large sizes when the retardation Mie effects become strong or dominant.

**Efficiency of a process**

First, we look at the electromagnetic properties of NRs. The corresponding parameter would be scattering efficiency,

$$e_{\text{scat}} = \frac{P_{\text{scat}}}{P_{\text{ext}}} = \frac{\sigma_{\text{scat}}(\omega)}{\sigma_{\text{ext}}(\omega)} \qquad (18)$$

Then, the whole picture of optoelectronic phenomena for NCs with various shape can be revealed by looking at energy efficiencies as was defined by us in Ref. [11]:

$$Eff_{\text{HEs}} = \frac{P_s}{P_{\text{ext}}} = \frac{Rate_{\text{HE}}}{(P_{\text{abs}} + P_{\text{scat}})/\hbar\omega}$$

$$Eff_{\text{High-E HEs}} = \frac{P_s}{P_{\text{ext}}} = \frac{Rate_{\text{High-E HE}}}{(P_{\text{abs}} + P_{\text{scat}})/\hbar\omega} \qquad (19)$$

$$Eff_{\text{d-holes}} = \frac{P_{\text{interband}}}{P_{\text{ext}}} = \frac{Rate_{\text{d-holes}}}{(P_{\text{abs}} + P_{\text{scat}})/\hbar\omega}$$

The efficiencies and the cross sections obey:

$$\sum_\gamma Eff_\gamma = 1$$

$$\sigma_{ext} = \sum_\gamma \sigma_\gamma \qquad (20)$$

where $\gamma$ the index of a process, i.e., $\gamma$ = scat, HEs, d-doles, and Drude. Then, a cross-section of a process is defined as

$$\sigma_\gamma = \frac{P_\gamma}{I_0} = \frac{Rate_\gamma / \hbar\omega}{I_0} \qquad (21)$$

We will soon observe that the efficiencies of generation of HEs, over-barrier HEs, and energetic d-holes depend on the material composition, geometry, and size. To understand how the energy efficiencies behave, we need to consider several factors:

- Electromagnetic and near-field effects
- Interband transitions
- Shape effects leading to the formation of hot spots
- A type of plasmon
- The spectral positions of plasmon peaks

Regarding the shape effects in plasmonic catalysis, we have found through comparison of nanocrystals of different shapes[31,32] that they play a crucial role. For example, it was shown recently that the gap-plasmon geometries with strong and extended hot spots featured significant enhancement in HE production.[33–35]

Another parameter that proves useful for assessing HEs and d-holes in the experiment is the material efficiency, which we defined in our previous work[31,32] to compare NCs of different shapes.

$$Eff_{HEs, material} = \frac{\sigma_{HE}}{V_{NC}}$$

$$Eff_{d\text{-}holes, material} = \frac{\sigma_{d\text{-}holes}}{V_{NC}} \qquad (22)$$

The units of this parameter are 1/nm, and it represents the efficiency of a plasmonic catalyst. This parameter is analogous to the mass extinction coefficient commonly used in aerosol optics.

Lastly, we can define **photothermal efficiency** as the ratio of all energy losses to total extinction.

$$Eff_T = \frac{P_{abs}}{P_{ext}} = \frac{P_{abs}}{P_{abs} + P_{scat}} \qquad (23)$$

**Au and Ag Nanorods.**

**Gold nanorods.** Figure 3 illustrates the exceptional tunability of the longitudinal plasmon resonances in Au NRs and Ag NRs.[36,37] From the first reports of the synthesis,[38–42] AuNRs attract much attention due to their optical properties caused by the presence of two plasmon resonances (transversal and longitudinal) and strong electromagnetic fields concentrated near the AuNRs tips. The position of the longitudinal LSPR depends linearly on the aspect ratio of GNRs, making the latter a tunable and versatile platform for applications in biomedical technologies, plasmon-enhanced spectroscopy, and optoelectronic devices (more details in section S3, supporting information).[40,43,44] We also observe the same linear trend for the extinction spectra with a gradual growth of scattering efficiency with length (Figure 3). However, for small Au NRs with aspect ratios (denoted later $e_{NR}$) less than 7, absorption continues to be the dominant factor.

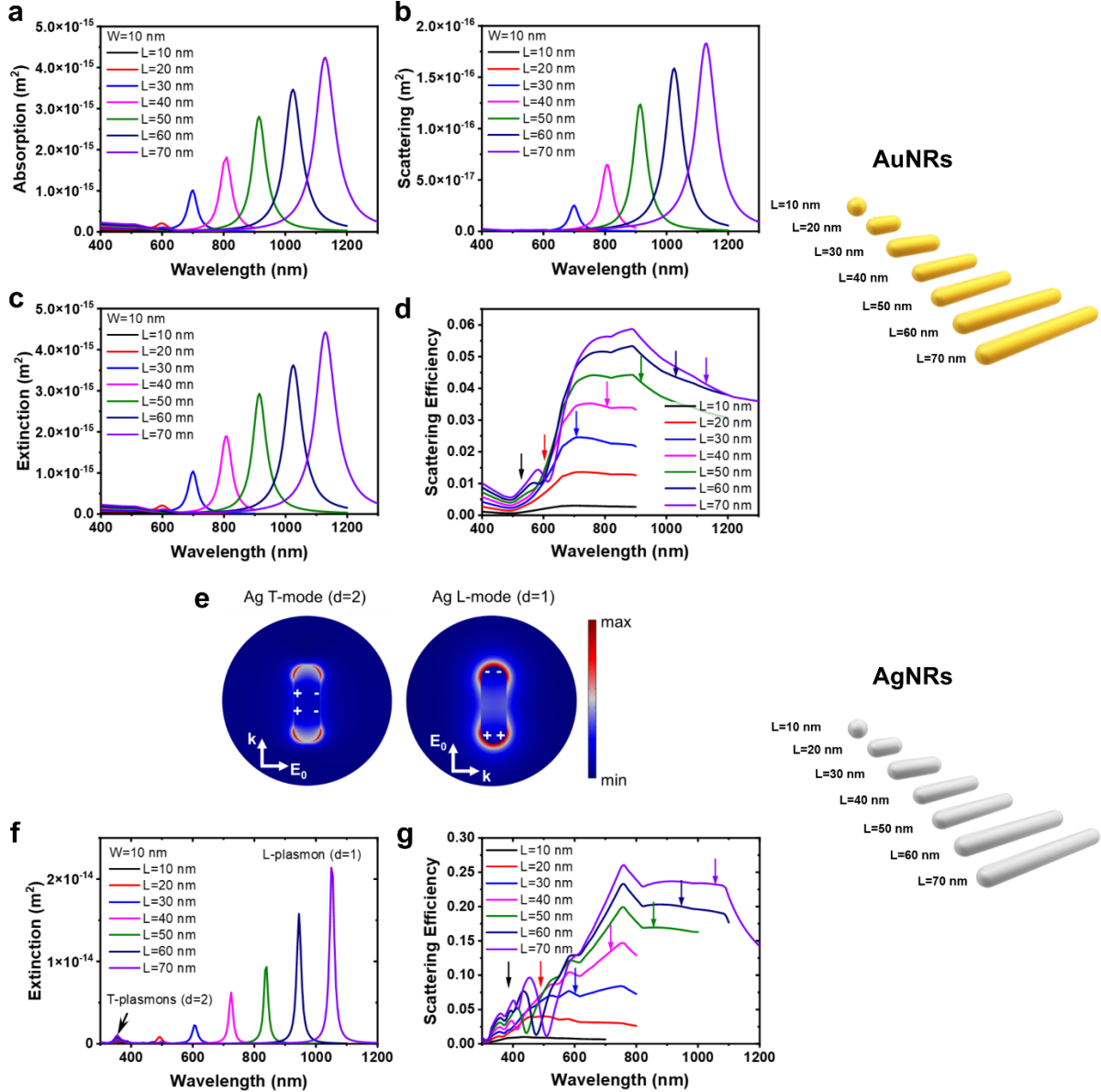

**Figure 3.** Electrodynamic responses of Au NRs in a solution (a-d). (a) Absorption, (b) scattering, and (c) extinction cross-sections of AuNRs of different sizes. (d) Scattering efficiency of AuNRs of different sizes. The arrows show the wavelengths of the plasmonic resonances. Electrodynamic responses of Ag NRs in a solution (e-g). The NR exhibits two major types of oscillations; d is the mode degeneracy. (e) E-field maps of T- and L-resonances of Ag nanorod. (f) Extinction cross-section of AgNRs of different sizes. (g) Scattering efficiency of AgNRs of different sizes.

In Figure 4, we show a complete set of plots for a small Au nanosphere (NSp) and an Au NR of $e_{NR} = 6$ (10 nm x 60 nm). For the NSp, the nonlinear equation for $\gamma_s$ becomes analytical, $\gamma_s \approx 3/2 \times (\hbar \cdot v_{F,Au} / a_{NSp})$, with $a_{NSp}$ being a NSp diameter. Overall, one can show that for a NC of arbitrary shape $\gamma_s = f_{NC}(\omega) \times (\hbar \cdot v_F / a_{eff})$, where $f_{NC}(\omega)$ and $a_{eff}$ are a NC-specific function and the effective size of a NC, respectively.[11] The function $f_{NC}(\omega)$ is numerical, in general. We observe in Figure 4a and 4e that the numerical nonlinear procedure for the Kreibig's parameter $\gamma_{s,n}$ converges fast, ensuring the energy conservation in the process of light absorption.

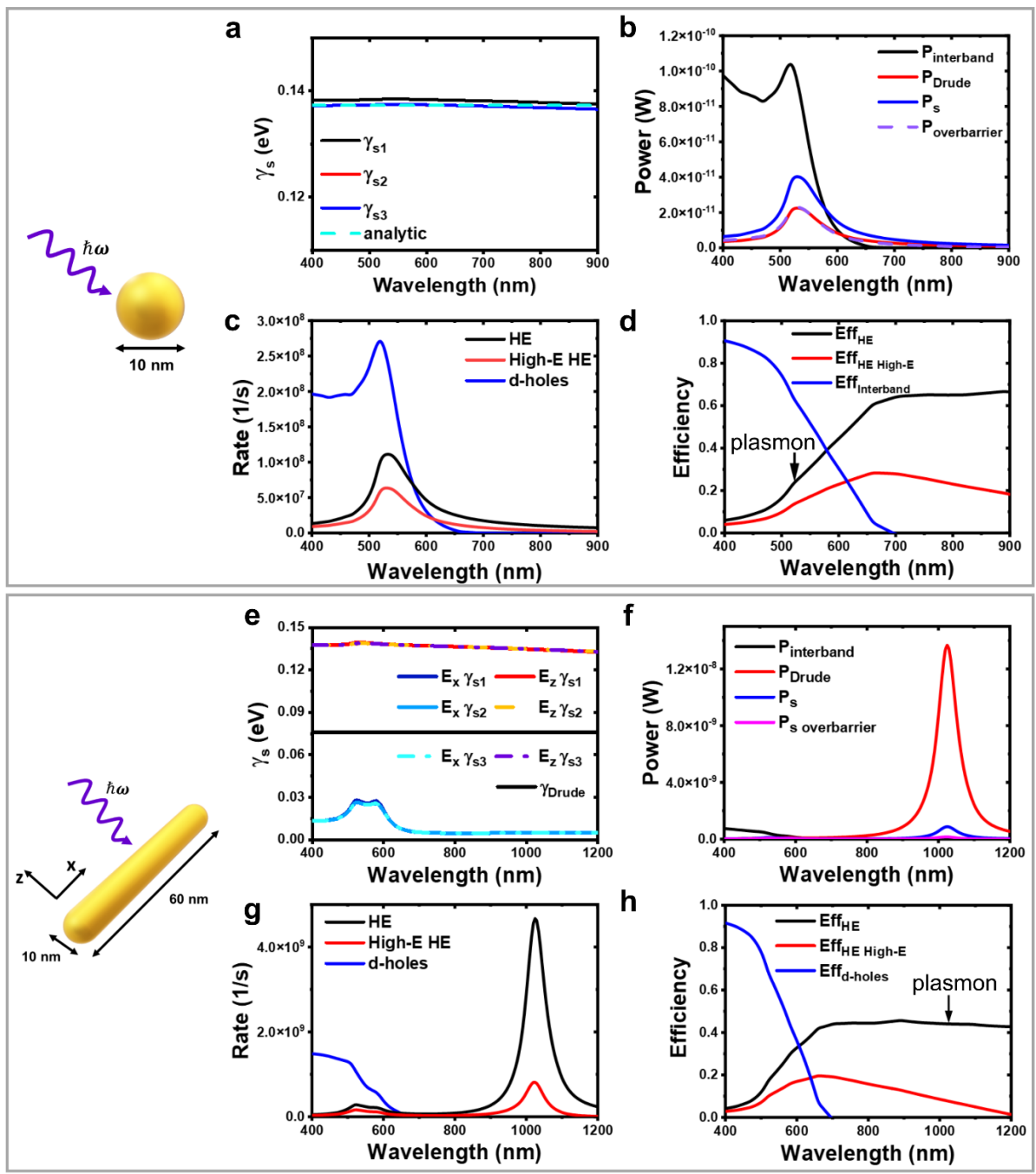

**Figure 4.** Energetics of Au NCs: nanosphere (a-d) vs. nanorod (e-h). These panels describe an Au NR with sizes 10 nm x 60 nm. The surface-scattering broadening parameter for NSp (a) and NR (e). Power of different dissipation mechanisms for NSp (b) and NR (f). Generation rates for the HEs, high-E HEs, and d-holes in NSp (c) and NR (g). Efficiencies of generation of excited carriers in NSp (d) and NR (h).

In both NCs in Figure 4, the generation of energetic d-holes is active for < 700 nm. The energy efficiency for the d-holes is very high in the blue, ~ 0.9 at 400 nm. This is because the interband transitions ($d \rightarrow sp$) dominate the absorption process in the blue and green regions. Although these transitions mainly contribute to the generation of energetic holes, in some cases they can also result in interband HE generation and further injection to the nearby molecule.[45] Also, combining the efficiencies of hot-carriers generation from interband and intraband transitions we can observe similar behavior to reported before.[45] Figure 4h also predicts the experimental results reported in [46] where it was shown that in AuNRs interband transitions more efficiently generate hot carriers as compared to the longitudinal LSPR. Regarding the intraband HEs, we see immediately the limitations – the generation happens in the continuum of states and, therefore, it will never be very efficient. For the NSp, $Eff_{HE} \sim 0.25$ at the plasmon peak (520 nm) and, for the Au NR, the HE efficiency increases to 0.4 at the L-plasmon resonance (1030 nm). In the latter case, the over-barrier efficiency is only 0.1 because of the Fowler-like factor in the rate, $(\hbar\omega - \Delta E_{bar})/\hbar\omega$. Overall, we observe that our formalism works well for this simple type of NC.

**Silver nanorods.** While colloidal Au NRs with high monodispersity have become a common staple in nanotechnology, synthesis, storage and applications of Ag NRs with small sizes are still challenging due to high oxidation rates in the presence of oxygen.[25,47,48] Also, the size and shape of the synthesized nanoparticles are harder to control compared to gold nanoparticles, and they need additional protective layers.[19,25,49] However, there were few successful reports on the Ag nanorods and nanobars photochemical[23] or polyol synthesis.[22,50] To overcome the Ag NRs synthesis limitations, researchers have discovered a brilliant approach: they fabricate Ag-coated Au NRs by starting with high-quality Au nanocrystals (more details in section S3, Supporting information).[16–19] Before studying properties of the core-shell NPs, we would like to compute and present purely silver NRs in this context (Figures 3i-g and 5).

Firstly, we observe that the L-plasmon peaks in the case of Ag are much sharper than those of Au NRs. Secondly, the scattering efficiency increases rapidly with the size of the NRs (Figure 3g), indicating that the electromagnetic Mie-like effects in the Ag-based systems are enhanced compared to the Au nanocrystals, as expected.

Figure 5 shows the energy structure of the Ag plasmons. First, we observe strong activity in the UV: at 385 nm for the spherical shape and at 359 nm for the NRs. Secondly, the intraband-HE efficiencies as compared to gold NCs are much higher. It is expected since Ag possesses more pronounced quantum effects because the mean free path of an electron in Ag is much longer and the mobile carriers forming the plasmon oscillation "feel" the surfaces stronger. In the UV resonance, the intraband HE efficiency gets up to ~ 50% in both NCs (Figure 5c,f). Simultaneously, for the red plasmon in the NR geometry, we obtain similar numbers of $Eff_{HE}$ ~ 50% .

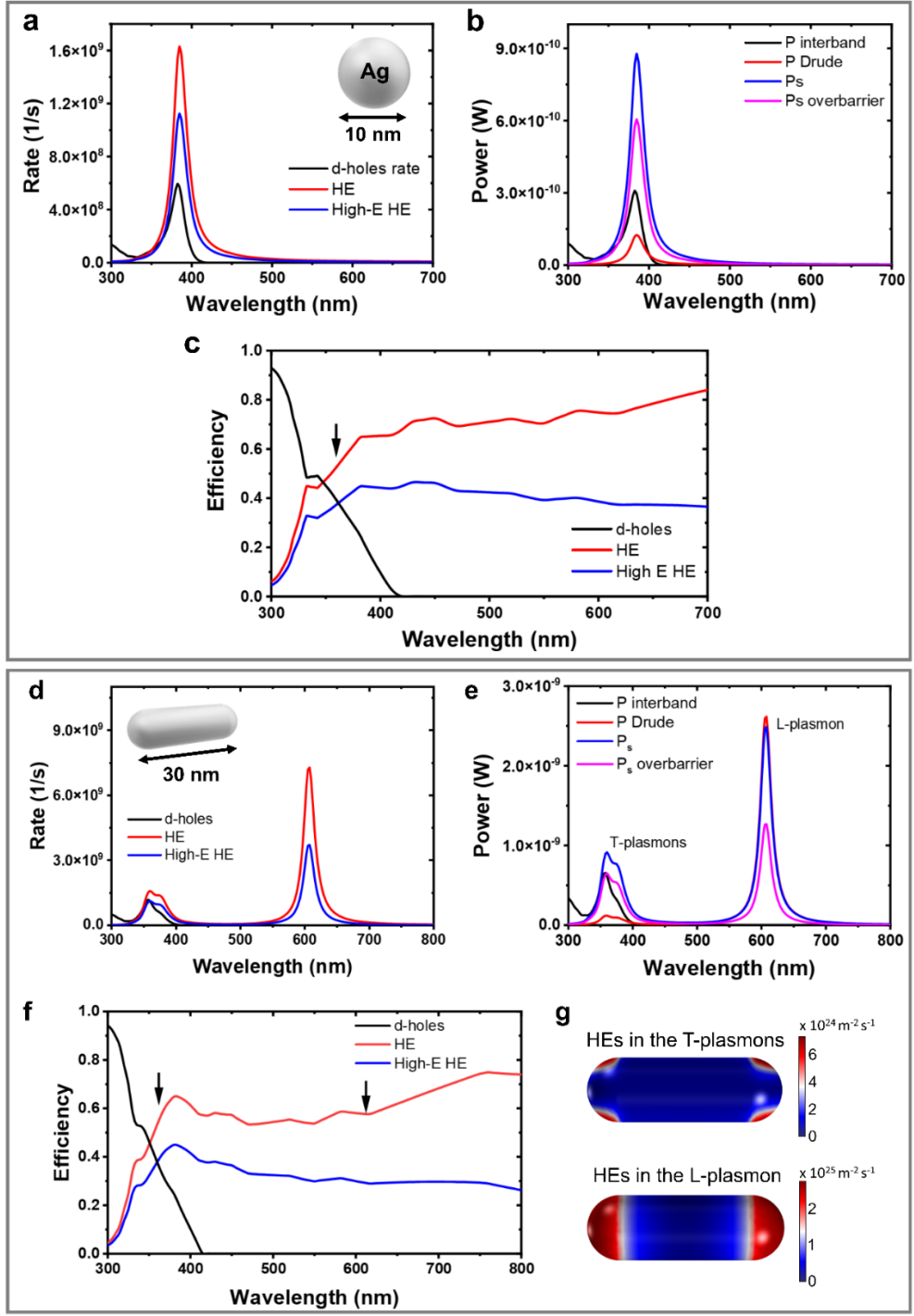

**Figure 5.** Energetics of Ag NCs in a solution. (a-c) Results for a spherical Ag NC and its physical structure of energy dissipation. (d-g) Data for an Ag NR in solution. Its spectra cover both UV and Vis spectral intervals. Generation rates for the HEs, high-E HEs, and d-holes in Ag NSp (a) and Ag NR (d). Power of different dissipation mechanisms for Ag NSp (b) an Ag NR (e). Efficiencies of generation of excited carriers in Ag NSp (c) and NR (f). Surface maps for the generation rates in resonances (g).

Strikingly, the UV plasmons contain mostly intraband HEs and interband d-holes with a minimal contribution of the Drude dissipation – in other words, mostly quantum transitions contribute to the dissipation process. At the same time for the L-plasmon, we observe a completely different picture. The HE-efficiency stays high (~ 50%), but the Drude dissipation becomes very active (nearly 50% in Figure 5e,f). This type of dissipation, the Drude processes, leads to the generation of low-energy excitations in the Fermi sea – the Drude mechanism of excitation occurs via the classical acceleration of an electron near the Fermi surface. To summarize, we see different types of dissipation for the UV and red plasmons, and we also observe a striking difference in the intraband-HE efficiencies for Au and Ag.

Figure 6 provides a summary for Au and Ag NRs. Higher HE efficiencies and stronger scattering effects for the Ag case come from the fact that Ag possessed high mobility of electrons (i.e., a lower value for the Drude broadening parameter). Therefore, quantum effects become much stronger in the case of Ag.

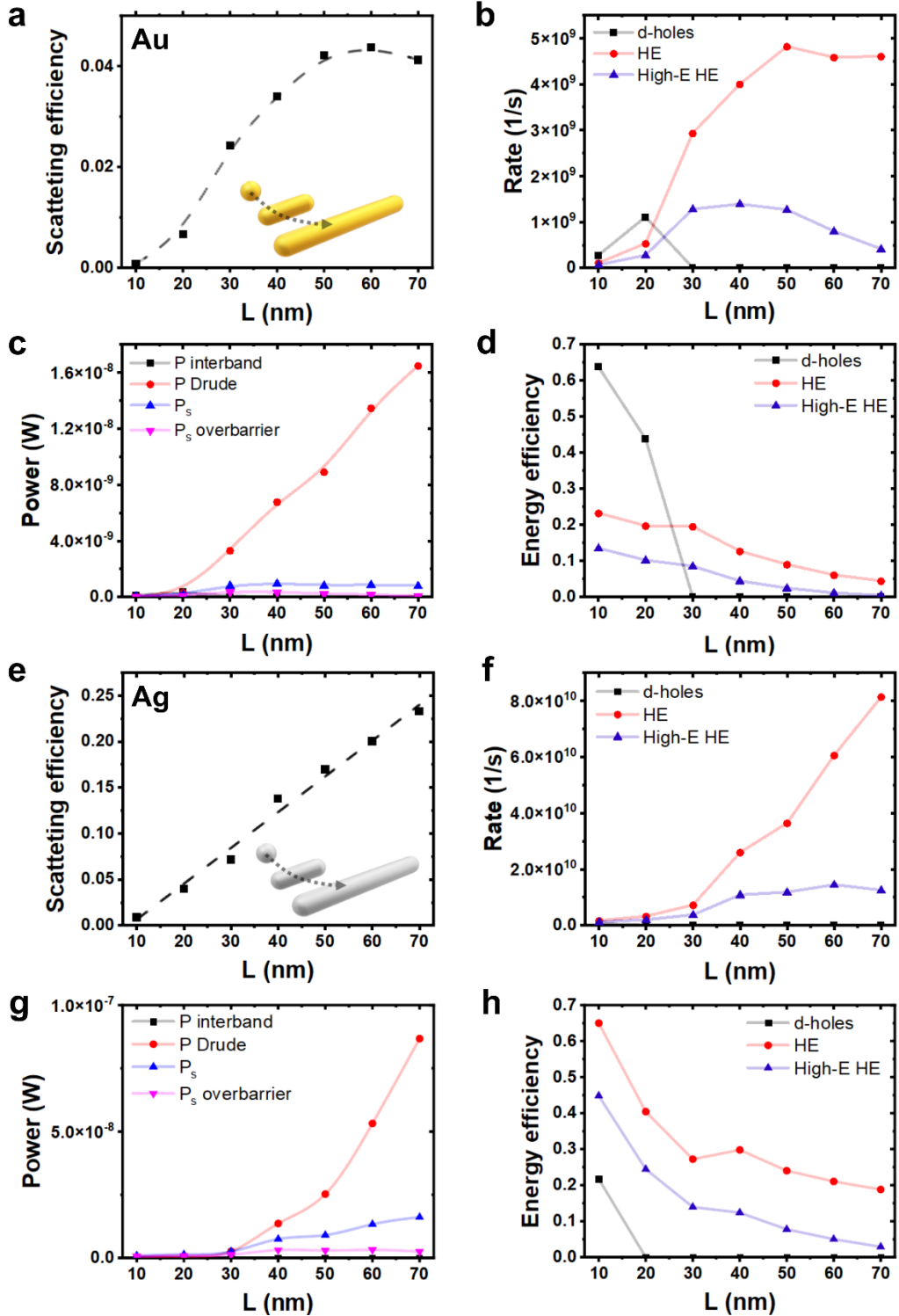

**Figure 6.** Summary for the efficiencies of scattering and dissipation process in noble-metal NRs in a solution. (a-d) Au NRs and (e-h) Ag NRs; W=10 nm and L=10-70 nm. Scattering efficiency in resonance for Au (a) and Ag (e) NRs. Generation rates for the HEs, high-E HEs, and d-holes in Au (b) and Ag (f) NRs. Power of different dissipation mechanisms for Au (c) an Ag NRs (g). Energy efficiencies of generation of excited carriers for Au (d) and Ag (h) NRs.

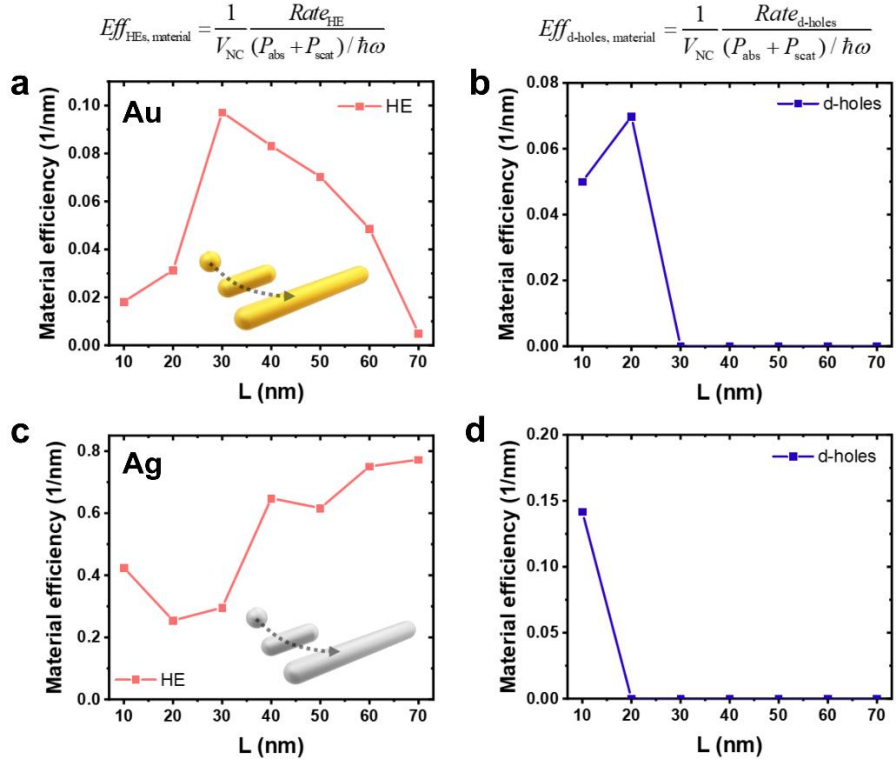

**Figure 7.** Assessing HEs and d-holes in monometallic NRs using the material efficiency parameter. (a,b) Au NRs and (c,d) Ag NRs with L=10-70nm, and W=10 nm. This figure of merit was proved very useful in our previous studies for comparing different shapes in a set of experiments.[31,32]

**Material parameter for Au & Ag.** Before coming to the more complex Au@Ag NRs, we look at the materials efficiencies of monometallic NRs (Figure 7). The HE generation is a quantum surface effect and one can expect the efficiency of the L-plasmon will decrease with the volume, $\sim 1/V \sim 1/L$. However, the behaviors in real material systems are more complex. For Au NRs in Figure 7a, $Eff_{HEs,\,material}(\lambda)$ increases till L ~ 30 nm since the L-plasmon resonance is gaining intensity for this interval, while moving away from the very intense interband transitions. Then, for L>30nm, the material efficiency declines since we have a decrease in the ratio $S_{Eff\,HE}/V \sim 1/L$, where $S_{Eff\,HE}$ is the an effective area where the HEs get generation; for the L-plasmon, the hot electrons are created at both ends, i.e., $S_{Eff\,HE} \sim 2\pi r^2$, where $r$ is the x-y cross-sectional radius of the NR. The decrease of $Eff_{HEs,\,material}(\lambda)$ for long L is faster in Figure 8a since the electromagnetic enhancement typically decreasing in long NRs.

For Ag NRs, the picture is different, as expected (Figure 7c). Here, the interband transitions are less important for the plasmon peak since they are mostly in the UV. In NR with a high-quality Drude metal (i.e., small $\gamma_D$), the L-plasmon peak intensity is increasing rapidly with L due to (a) increasing volume and (b) higher field-amplification factors inside a NR (Figure 3f). The extinction peak rises almost like $L^2$ (Figure 3f). Therefore, HE production tends to increase with L too, however the growth of the efficiency with L in Figure 7c is not that strong. For longer Ag NRs, we expect a decrease of the material intraband-HE efficiency since the electromagnetic effects and scattering will boost rapidly in long NRs. Here, we do not compute that spectral region since we focus on the colloidal NCs, which are typically not so long. For this Mie-regime (long L), we would need lithographic nanowires.

The d-holes behavior is the last point of this section. Of course, the d-hole materials efficiency at the plasmon resonance will vanish for $\lambda > \lambda_{\text{interband cutoff}}$ (690 nm for Au and 420 nm for Ag). This is what we observe in Figures 7b,d.

**Au@Ag core-shell NRs: the 2-$\gamma_s$ formalism.**

Next, we describe the properties of the hybrid core-shell nanorods, which are the most interesting and nontrivial. The core-shell configuration also proved to be more chemically and morphologically stable and have narrower plasmon linewidths than pure Ag NRs while requiring similar stabilizing agents as gold nanoparticles.[25,49,51] Already in our previous publication,[18] we observed an intriguing behavior of extinction and hot electrons in the UV – the Au@Ag core-shell NRs possess strong UV lines at about 350 nm. Such energetic excitations are highly desirable for photocatalysis and may enable realization of some of the challenging reactions like C-based fuels' synthesis. Our calculations explain this very unusual activity in the UV, involving high generation rates of HEs and d-holes in the UV region. However, we need to go step by step, and we start with the extinctions (Figures 8 and 9).

In Figures 8a,b, we show separately the UV- and Vis-intervals. In the UV interval, we see the changes of the UV plasmon in the Au@Ag Sph and NR geometry. The UV plasmon of the hybrid sphere at 384 nm is dipolar mode of a NC (Figure 8c). When a Sph NC develops into a NR, the UV peak transforms remarkably (Figure 8a) – it shifts deeper to the UV and reappears at 351 nm. The peak stays sharp in the Au@Ag hybrid despite the strong interband transient in Au. Why? By looking at the field map on Figure 8c, we figure out that this T-plasmon is confined in the Ag-shell, its fields do not propagate much to the Au core. It resembles a cylindrical capacitor (Figure 8c). Therefore, the UV T-plasmon at ~ 350 nm does not experience strong interband damping in Au. However, Ag itself also has interband transitions, which certainly contribute to the UV T-plasmon peak. This gap-plasmon peak is less sharp than that in a monometallic Ag Sph. Figure 8d shows an evolution of the wavelength of the T-plasmon. Figures 8b,e,j,i focus on the L-resonances which is a longitudinal dipolar plasmon in the Au@Ag NR and has two hot spots at the ends.

The plasmonic peaks in the HE production in the UV and Vis spectral ranges follow the extinction (Figures 8f,g and Figure 9), however the behavior of their magnitude/shape are, in general, different to those of the optical cross-sections. Of course, the Vis-NIR plasmon is the most intense in the spectra on Figure 9 since it has a huge dipole moment. However, the HEs and d-holes in the UV carry much more energy, which can be used in photoreactions. The dependence of the HE rates in the UV with L is approximately linear since the HE generation takes place mainly at the external side surfaces of a NR:

$$Rate_{HE} = \frac{1}{4} \times \frac{2}{\pi^2} \times \frac{e^2 E_F^2}{\hbar} \frac{1}{(\hbar\omega)^3} \int_{S_{NC}} |E_{\omega, normal}(\theta,\varphi)|^2 ds \propto S_{eff} \frac{1}{(\hbar\omega)^3} f_{enh}(\omega)$$

where $S_{eff} \sim L \times \pi r^2$ is the area of the side surfaces and $f_{enh}(\omega)$ is the field enhancement factor for the T-plasmon.

The T-plasmon in the UV is also active regarding the d-hole excitations in Ag – such a process also leads to a strong peak at ~ 350 nm (Figure 8h). At the same time, the L-plasmon is not able to excite the interband d-sp transitions in Ag, and its hot-carrier production is limited to the intraband HE generation.

We also observe a fundamental difference between the T- and L-plasmons. The L-plasmon suffers from the electromagnetic decay in the NIR – we see that the scattering efficiency for the L-mode becomes > 0.5 for the longest NRs (Figure 9c, g). The scattering leads to additional mechanisms of energy dissipation and to the broadening of a plasmon peak. Therefore, this mechanism will tend to reduce the HE efficiency in the red and NIR intervals. Simultaneously, the T-plasmon in the Ag shell has a small dipole and shows very weak scattering. This provides another advantage of the UV plasmon for the HE generation, along with the ultra-high energy of the UV HEs.

Important to note that in this section we focused on the generation of hot carriers in the distinct bands of L-plasmon, T-plasmon, and interband and did not pay attention to special shapes/regimes when those excitations can interact creating interesting Fano-like spectral shapes.[52,53]

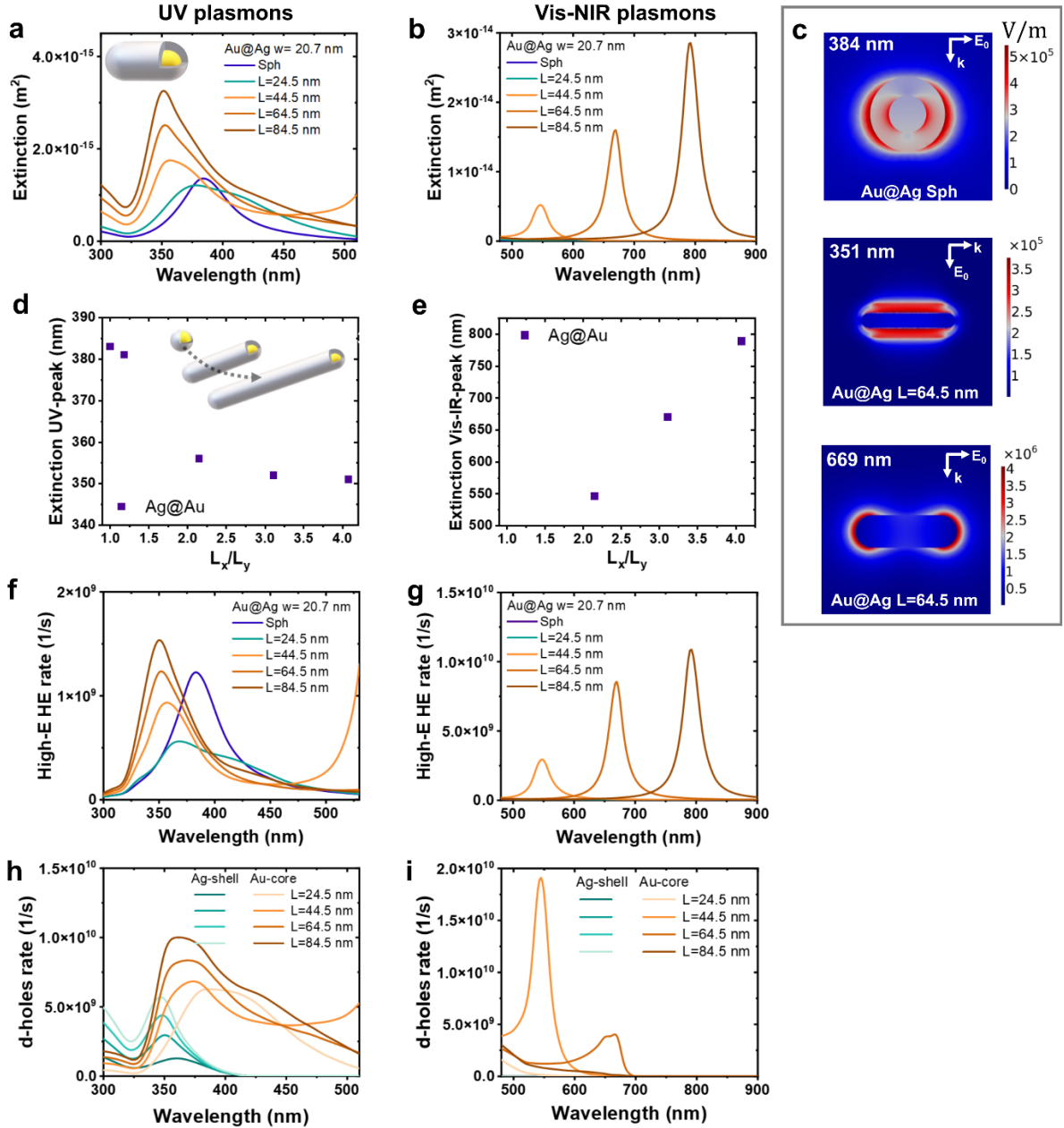

**Figure 8.** Plasmonic properties of the Au@Ag core-shell NRs. In this figure, we show the UV and Vis intervals separately, for convenience. (a-b) Extinction cross-sections of Au@Ag NRs. (c) E-field maps of the Au@Ag NRs resonances. (d-e) Dependences of the extinction peak positions on aspect ratio of the NRs. (f-i) Energetic properties in terms of the rates of generation: generation rates for the high-E HEs (f,g) and d-holes (h,i).

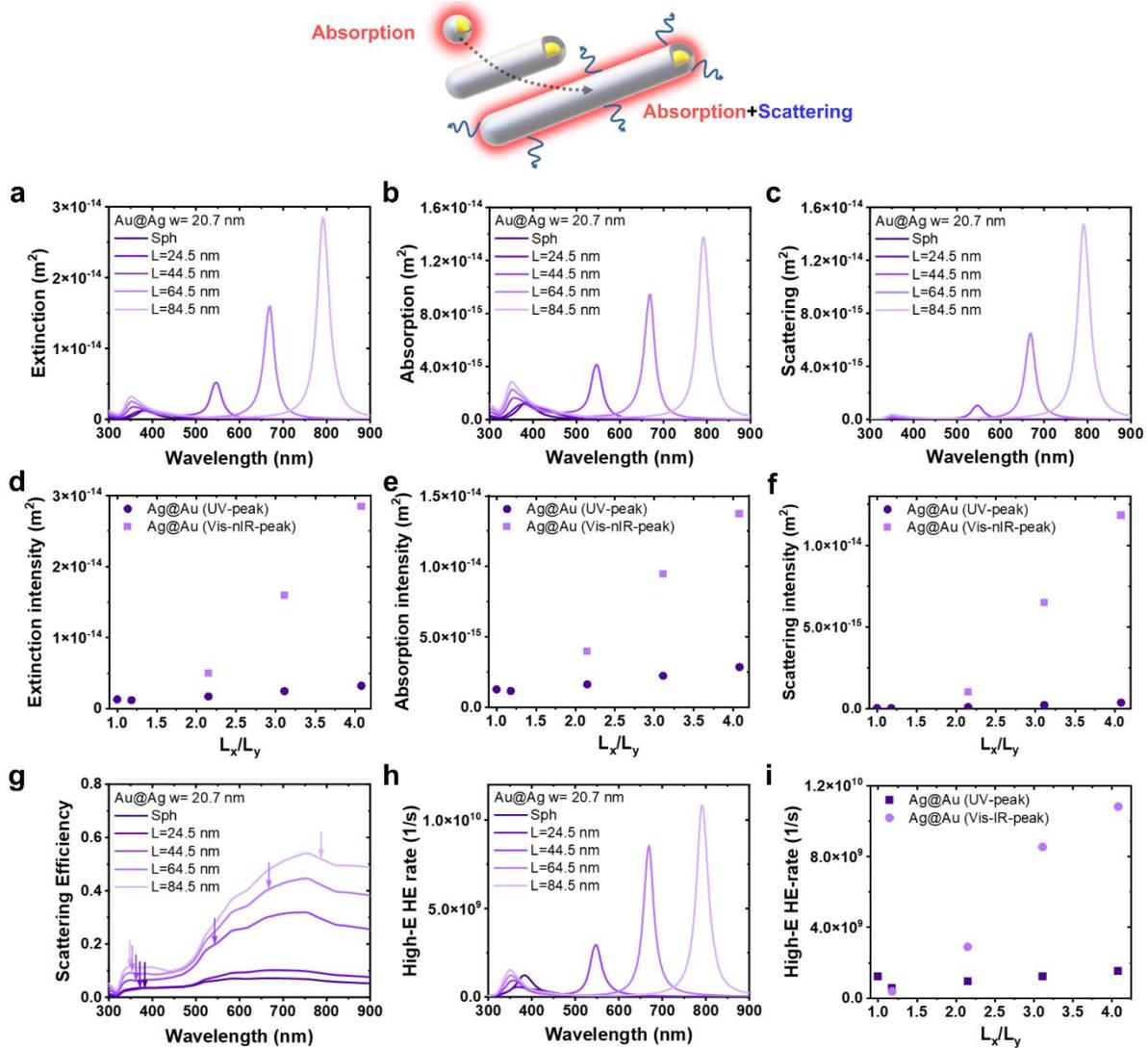

**Figure 9.** Energetic and optical properties of the Au@Ag core-shell NRs in the whole, broadband interval. Extinction (a), absorption (b), and scattering (c) cross-sections of Au@Ag NRs. (d-f) Dependences of the extinction (d), absorption (e), scattering (f) intensity on aspect ratio of the Au@Ag NRs. Scattering efficiency (g) and energetic properties in terms of the rates of generation (h) of the Au@Ag NRs. (h) Dependence of the electronic properties at plasmonic resonances on the aspect ratio of the Au@Ag NRs.

The HE generation efficiencies, which are computed in this study, correspond to the decay parameter $\gamma_s$ that was used in a number of experimental papers.[54] We show the internal efficiencies ($Eff$), while the experimental papers report external efficiencies, $Eff_{external}$, which is < or << $Eff$. In the SI S2, we give model for the HE transfers and a few examples of experimental efficiencies. We should note that the understanding the internal efficiencies are crucial since it tells us what high-energy carriers are created (intraband HEs, interband holes or Drude electrons) and in what rate. It is well established that the limitation factors for a metal absorber are the short lifetime and transfer time of high energy carriers, although HE in a metal nanocrystal are generated intensively near the surface and are also moving fast.[11,55] The following review and research papers [11,54,55] can be used to read more on the non-thermalized (HEs and d-holes) and thermalized (Drude electrons) carriers under both CW and pulsed excitations.

**Conclusions**

We summarize our findings in Figure 10. The uniqueness of the Au@Ag core-shell NRs is in their broadband spectra extending from the UV to the NIR. These NCs possess the unique property to create intensive peaks in the UV at about 350 nm, which are the slab-like plasmons in the Ag shell.[17,18] When we compare with the other examples of plasmon peaks within UV plasmonics, like in the Al NCs[56–59] for instance, we see that the UV plasmons in the high-quality and monodisperse Au@Ag systems are sharper and also appear further in the UV.

Regarding the efficiencies in Figure 10, their behaviors are very characteristic, reflecting the physics of the electromagnetic and relaxation processes in a NC with a specially designed shape. The energy and material efficiencies of hot-carrier generation depend on the following factors:

- The surface-to-volume ratio in NC. The rate of the surface-assisted HE generation (i.e., Kreibig's mechanism) is proportional to the surface area where the HEs are actively generated. In contrast, the Drude dissipation rate is proportional to the total volume. Therefore, the ratio of $S_{eff}/V$ is involved. The energy rates decrease with the size generally as $1/L_{eff}$, assuming a Drude metal (no interband transitions) and a small size of NC.
- Interband transitions in the metals are crucial for the UV and Vis plasmons. In the Au@Ag core-shell material, the slap-like plasmon in the Ag shell remains sharp since the interband process in Au is not highly active at 350 nm yet. In Au, the interband processes govern the absorption below 600 nm and are being used in photochemistry for the d-hole generation.

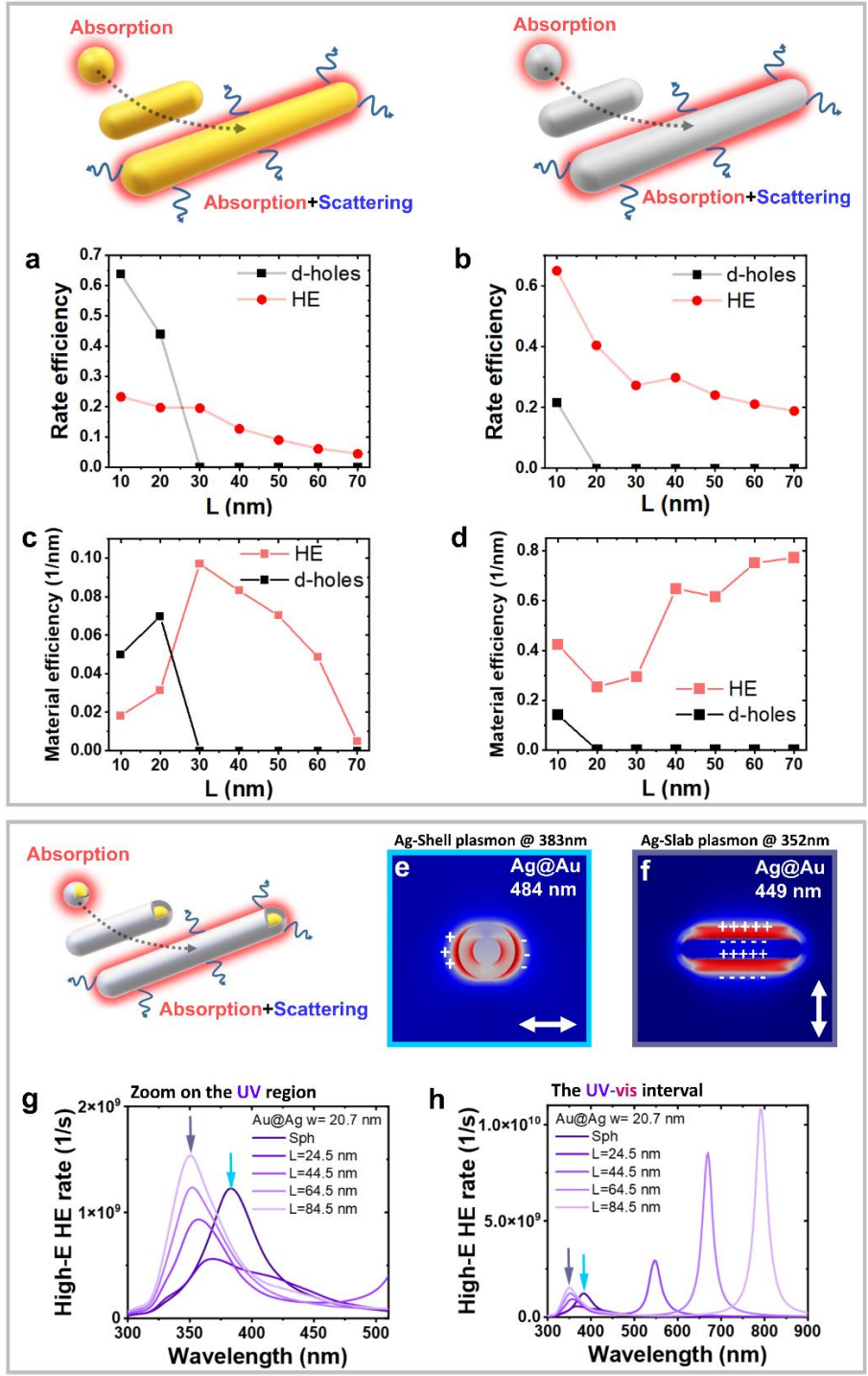

**Figure 10.** Summary of the hot-carrier and electromagnetic properties of Au, Ag and Au@Ag NRs. The efficiency parameters are shown at the main plasmonic peak in the Vis-NIR interval. (a-d) Monometallic NRs and their nontrivial behaviors. Energy efficiencies of generation of excited carriers for Au (a) and Ag (b) NRs. Material efficiencies of generation of excited carriers for Au

(c) and Ag (d) NRs. (e,f) E-field maps of the Au@Ag NRs resonances. (g,h) HE generation rates for the UV and visible plasmons in the bimetallic core-shell NRs.

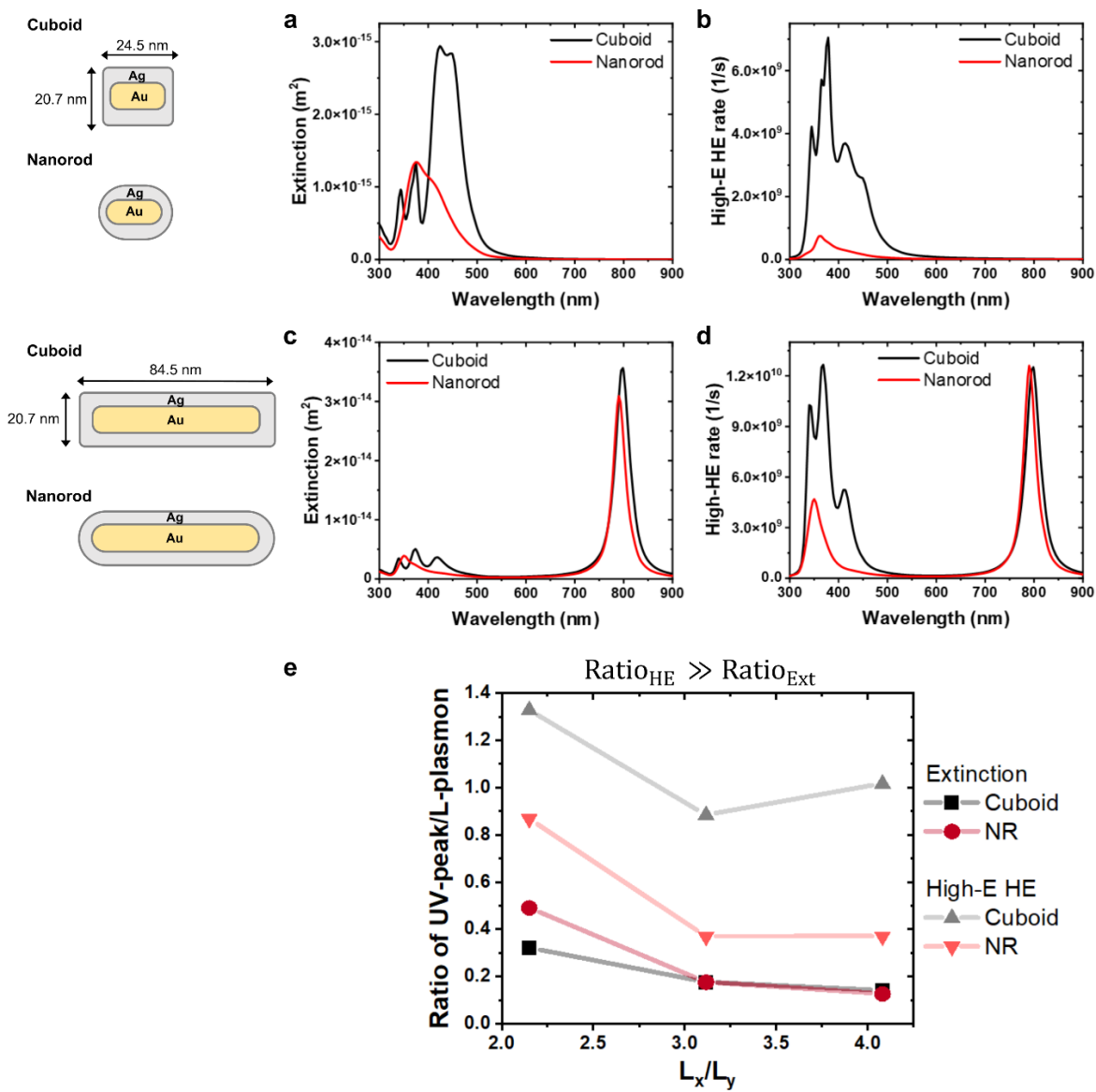

**Figure 11.** Electrodynamic responses of Au@Ag nanorods and cuboids. Extinction (a) and generation rates for the high-E HEs (b) of Au@Ag nanorods and cuboids with L=24.5 nm and W=20.7 nm. Extinction (c) and generation rates for the high-E HEs (d) of Au@Ag nanorods and cuboids with L=84.5 nm and W=20.7 nm. (e) Ratio of UV-peak/L-plasmon intensity in extinction and high-E HE generation rates for nanorods and cuboids.

- Electromagnetic effects start playing an essential role in the long and large NRs – those effects reduce the field enhancement and create more scattering. Therefore, the electromagnetic Mie-like effects work against achieving high efficiencies.
- The UV plasmon resonances are unique and strong in the Au@Ag systems. Those plasmons are T-modes and do not suffer from electromagnetic decay. The T-resonance in the Ag shell at 350 nm generates both high-energy HEs and energetic d-holes.
- The hot-carrier generation efficiencies can be further optimized by modulation of the shapes and surface-to-volume ratio in NC. In more cuboidal-like shapes (Figure 11, Figures S6-S8, Table S2) which are often observed in experiments[18] the generation of the HEs (Figure 11b,d and Figure S8g) in UV spectral range is largely augmented. Interestingly, the changes in extinction ratio between visible and UV plasmon resonances do not fully reflect the changes in the hot-carriers generation rates (Figure 11e). For cuboids, the High-E HE generation rates of the UV plasmon resonances reach and overcome the rates of the L-plasmon.

Where are the potential uses of the plasmonic HEs and hot holes? – in sensing, nanotechnology, photocatalysis, and electrophotochemistry. For instance, hot electrons and holes in NC are proven to be a fantastic mechanism for photo-growth employed in colloidal nanotechnology.[60] Plasmons and plasmonic hot spots generate energetic carries driving reactions locally and creating new shapes. One impressive recent example is the re-shaping of nanostructures using chiral light.[61,62]

(61) Saito, K.; Tatsuma, T. Chiral Plasmonic Nanostructures Fabricated by Circularly Polarized Light. *Nano Letters* **2018**, *18* (5), 3209–3212. https://doi.org/10.1021/acs.nanolett.8b00929.

(62) Ávalos-Ovando, O.; Santiago, E. Y.; Movsesyan, A.; Kong, X.-T.; Yu, P.; Besteiro, L. V.; Khorashad, L. K.; Okamoto, H.; Slocik, J. M.; Correa-Duarte, M. A.; Comesaña-Hermo, M.; Liedl, T.; Wang, Z.; Markovich, G.; Burger, S.; Govorov, A. O. Chiral Bioinspired Plasmonics: A Paradigm Shift for Optical Activity and Photochemistry. *ACS Photonics* **2022**, *9* (7), 2219–2236. https://doi.org/10.1021/acsphotonics.2c00445.32

# Supporting Information
## 1. Formalisms
### 1.1 Bi-metallic nanorods.

According to the surface-scattering formalism developed in Refs [1–4], the integrals involved in the $\gamma_s$-model of a metal NC are taken in the following ways:

$Lim_{\varepsilon \to 0}$:

$$F_{V,Au,n} = \frac{\int_{V_{Au}} dV\, \mathbf{E}_{\omega,n} \cdot \mathbf{E}^*_{\omega,n}}{V_{Au}}$$

$$F_{V,Ag,n} = \frac{\int_{V_{Ag}} dV\, \mathbf{E}_{\omega,n} \cdot \mathbf{E}^*_{\omega,n}}{V_{Ag}}$$

$$F_{S-\varepsilon-\text{internal},Au,n} = \frac{\int_{S_{Au}} ds\, |E_{\text{normal, n}}(\theta,\varphi)|^2}{S_{Au}} \quad \text{- inside Au close to the interface}$$

$$F_{S+\varepsilon-\text{internal},Ag,n} = \frac{\int_{S_{Ag}} ds\, |E_{\text{normal, n}}(\theta,\varphi)|^2}{S_{Ag}} \quad \text{- inside Ag close to the interface}$$

$$F_{S-\text{external},Ag,n} = \frac{\int_{S_{Ag}} ds\, |E_{\text{normal, n}}(\theta,\varphi)|^2}{S_{Ag}}$$

The integration surfaces are defined as:

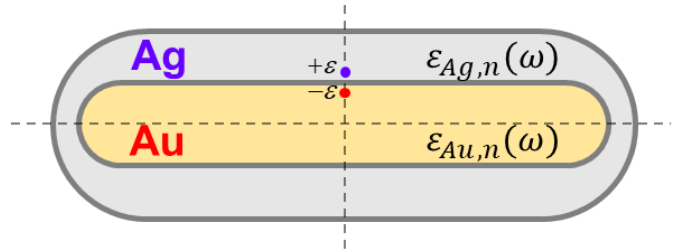

**Figure S1.** Schematics of the integration surfaces of the Au@Ag nanorod

The nonlinear theory of Refs. [3,4] come from the energy conservation equations:

The dissipation due to the Kreibig's parameter $\gamma_{s,Au}$ = The dissipation due to the quantum mechanism of HE generation

Here the Kreibig's dissipation is a contribution due to HEs computed using the bulk epsilon ($\varepsilon_{metal,s}(\omega)$) in one metallic component, whereas the dissipation due to the quantum HE generation is given by the rate of HE generation computed from quantum mechanics near the surfaces,[1] i.e., by the equations $P_{abs,HE,Au} = Rate_{HE,Au} \times \hbar\omega$ and $P_{abs,HE,Ag} = Rate_{HE,Ag} \times \hbar\omega$.

$$P_{abs,HE,Au} = \left\langle \int_{Au} dV\, \mathbf{j}_S \cdot \mathbf{E} \right\rangle_{time} = \text{Im}[\Delta\varepsilon_{Au,s}] \cdot \varepsilon_0 \frac{\omega}{2} \int_{Au} dV\, \mathbf{E}_\omega \cdot \mathbf{E}_\omega^* \equiv Rate_{HE,Au} \times \hbar\omega$$

$$\text{Im}[\Delta\varepsilon_{Au,s}] \approx \frac{\omega_p^2}{\omega^3} \gamma_{s,Au}$$

$$\Delta\varepsilon_{Au,s} = \varepsilon_{Au,s} - \varepsilon_{Au,bulk}$$

(S1)

$$P_{abs,HE,Au} = \left\langle \int_{Au} dV\, \mathbf{j}_S \cdot \mathbf{E} \right\rangle_{time} = \text{Im}[\Delta\varepsilon_{Au,s}] \cdot \varepsilon_0 \frac{\omega}{2} \int_{Au} dV\, \mathbf{E}_\omega \cdot \mathbf{E}_\omega^* \equiv Rate_{HE,Au} \times \hbar\omega$$

$$\text{Im}[\Delta\varepsilon_{Au,s}] \approx \frac{\omega_p^2}{\omega^3} (\gamma_{s,Au})$$

$$\Delta\varepsilon_{Au,s} = \varepsilon_{Au,s} - \varepsilon_{Au,bulk}$$

In our model the HE generation occurs at the interfaces:

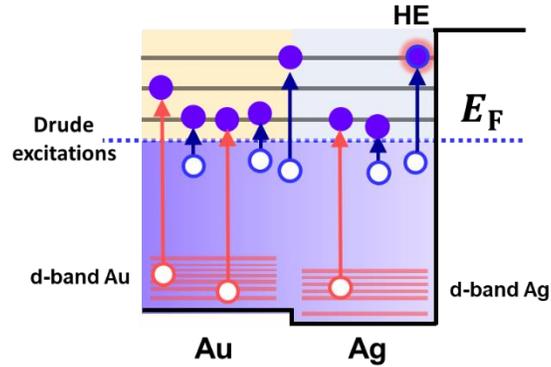

Figure S2. Hot-e generation in the sp-bands of Au and Ag in Au@Ag nanocrystals.

The analytical equations for the rates of HEs are:

**Au-Ag interface:**

$Rate_{HE,Au, \text{at the Au-Ag interface}} =$

$$= \frac{1}{2} \times \frac{|\Delta E_F|}{\bar{E}_F} \left[ \frac{1}{4} \times \frac{2}{\pi^2} \times \frac{e^2 E_{F,Au}^2}{\hbar} \frac{1}{(\hbar\omega)^3} \int_{S_{Au-Ag}-\varepsilon} |E_{\omega,normal}(\theta,\varphi)|^2 ds + \frac{1}{4} \times \frac{2}{\pi^2} \times \frac{e^2 E_{F,Ag}^2}{\hbar} \frac{1}{(\hbar\omega)^3} \int_{S_{Au-Ag}+\varepsilon} |E_{\omega,normal}(\theta,\varphi)|^2 ds \right]$$

$$Rate_{\text{HE,Ag, at the Au-Ag interface}} = Rate_{\text{HE,Au, at the Au-Ag interface}} \quad \text{(S2)}$$

**Ag-matrix interface:**

$$Rate_{\text{HE,Ag, surafce}} = \frac{1}{4} \times \frac{2}{\pi^2} \times \frac{e^2 E_{\text{F,Au}}^2}{\hbar} \frac{1}{(\hbar\omega)^3} \int_{S_{\text{Ag}-\varepsilon}} |E_{\omega,\text{normal}}(\theta,\varphi)|^2 ds \quad \text{(S3)}$$

For some of the notations and parameters used above:

$$\Delta E_{\text{F}} = E_{\text{F,Au}} - E_{\text{F,Ag}}$$

$$\bar{E}_{\text{F}} = (E_{\text{F,Au}} + E_{\text{F,Ag}})/2$$

(S4)

Table S1. Materials parameters

| Parameters | Au | Ag |
|---|---|---|
| $\omega_p$ | 8.9 eV | 9.3 eV |
| $\gamma_{\text{D}}$ | 0.076 eV | 0.02 eV |
| $n_e$ | $5.74 \ast 10^{28}$ 1/m$^3$ | $6.27 \cdot 10^{28}$ 1/m$^3$ |
| $E_{\text{F}}$ | 5.43 eV | 5.75 eV |
| Work function | 4.6 eV | 4.6 eV |
| $|\Delta E_{\text{F}}|$ | 0.32 eV | |

The following assumptions were made in S2-S3:

(1) The HE generation in the sp-bands of Au and Ag occurs near the potential discontinues – see Fig. S2.

(2) The HE generated at the Au-Ag interface are divided equally between Au and Ag.

(3) $|\Delta E_{\text{F}}| \ll \bar{E}_{\text{F}}$. Indeed, $|\Delta E_{\text{F}}| = 0.32$ eV and $\bar{E}_{\text{F}} = 5.59$ eV. Therefore, $Rate_{\text{HE,Ag, surafce}} \gg (Rate_{\text{HE,Au, at ...}} + Rate_{\text{HE,Ag, at ...}})$.

(4) Finally, $\omega^2 \gg (\gamma_{\text{D}} + \gamma_s)^2$. This leads to $\text{Im}[\Delta\varepsilon_{\beta,\text{s}}] \approx \frac{\omega_{\text{p},\beta}^2}{\omega^3}(\gamma_{s,\beta})$, where $\beta = Au, Ag$ and $\Delta\varepsilon_{\beta,\text{s}} = \varepsilon_{\beta,\text{s}} - \varepsilon_{\beta,\text{bulk}}$.

Next, using (S1), we obtain the final set of equations.

$$\gamma_{s,Au} = \frac{|\Delta E_F|}{\bar{E}_F} \frac{1}{2} \left( \frac{3}{4} \times v_{F(Au)} \frac{S_{Au}}{V_{Au}} \frac{F_{S,Au-\varepsilon}(\gamma_{s,Au,n=0}=0, \gamma_{s,Ag,n=0}=0)}{F_{V,Au}(\gamma_{s,Au,n=0}=0, \gamma_{s,Ag,n=0}=0)} + \frac{3}{4} \times v_{F(Ag)} \frac{S_{Ag}}{V_{Ag}} \frac{F_{S,Ag+\varepsilon}(\gamma_{s,Au,n=0}=0, \gamma_{s,Ag,n=0}=0)}{F_V(\gamma_{s,Au,n=0}=0, \gamma_{s,Ag,n=0}=0)} \right) > 0$$

(S5)

$$\gamma_{s,Ag} = \frac{3}{4} \times v_{F(Ag)} \frac{S_{Ag}}{V_{Ag}} \frac{F_{S,Ag,ext}(\gamma_{s,Au,n=0}=0, \gamma_{s,Ag,})}{F_V(\gamma_{s,Au,n=1}=0, \gamma_{s,Ag,n=0}=0)} +$$

$$+ \frac{|\Delta E_F|}{\bar{E}_F} \frac{1}{2} \left( \frac{3}{4} \times v_{F(Au)} \frac{S_{Au}}{V_{Au}} \frac{F_{S,Au-\varepsilon}(\gamma_{s,Au,n=0}=0, \gamma_{s,Ag,n=0}=0)}{F_{V,Au}(\gamma_{s,Au,n=0}=0, \gamma_{s,Ag,n=0}=0)} + \frac{3}{4} \times v_{F(Ag)} \frac{S_{Ag}}{V_{Ag}} \frac{F_{S,Ag+\varepsilon}(\gamma_{s,Au,n=0}=0, \gamma_{s,Ag,n=0}=0)}{F_{V,Ag}(\gamma_{s,Au,n=0}=0, \gamma_{s,Ag,n=0}=0)} \right) > 0$$

It is a complete set of nonlinear integrodifferential equations having a unique solution for small $\gamma_{s,\beta}$. The way to solve it for a complex geometry and large sizes is to compute these equations numerically using the FEM (COMSOL Multiphysics is one good option) and the iteration procedure introduced by us in Refs. [3,4]:

$\gamma_{s,Au,n=0} = 0$

$\gamma_{s,Ag,n=0} = 0$

$\downarrow$

$$\gamma_{s,Au,n=1} = \frac{|\Delta E_F|}{\bar{E}_F} \frac{1}{2} \left( \frac{3}{4} \times v_{F(Au)} \frac{S_{Au}}{V_{Au}} \frac{F_{S,Au-\varepsilon}(\gamma_{s,Au,n=0}=0, \gamma_{s,Ag,n=0}=0)}{F_{V,Au}(\gamma_{s,Au,n=0}=0, \gamma_{s,Ag,n=0}=0)} + \frac{3}{4} \times v_{F(Ag)} \frac{S_{Ag}}{V_{Ag}} \frac{F_{S,Ag+\varepsilon}(\gamma_{s,Au,n=0}=0, \gamma_{s,Ag,n=0}=0)}{F_V(\gamma_{s,Au,n=0}=0, \gamma_{s,Ag,n=0}=0)} \right) > 0$$

$$\gamma_{s,Ag,n=1} = \frac{3}{4} \times v_{F(Ag)} \frac{S_{Ag}}{V_{Ag}} \frac{F_{S,Ag,ext}(\gamma_{s,Au,n=0}=0, \gamma_{s,Ag,n=0}=0)}{F_V(\gamma_{s,Au,n=1}=0, \gamma_{s,Ag,n=0}=0)} +$$

$$+ \frac{|\Delta E_F|}{\bar{E}_F} \frac{1}{2} \left( \frac{3}{4} \times v_{F(Au)} \frac{S_{Au}}{V_{Au}} \frac{F_{S,Au-\varepsilon}(\gamma_{s,Au,n=0}=0, \gamma_{s,Ag,n=0}=0)}{F_{V,Au}(\gamma_{s,Au,n=0}=0, \gamma_{s,Ag,n=0}=0)} + \frac{3}{4} \times v_{F(Ag)} \frac{S_{Ag}}{V_{Ag}} \frac{F_{S,Ag+\varepsilon}(\gamma_{s,Au,n=0}=0, \gamma_{s,Ag,n=0}=0)}{F_{V,Ag}(\gamma_{s,Au,n=0}=0, \gamma_{s,Ag,n=0}=0)} \right) > 0$$

$\downarrow$

$$\gamma_{s,Au,n=2} = \frac{|\Delta E_F|}{\bar{E}_F} \frac{1}{2} \left( \frac{3}{4} \times v_{F(Au)} \frac{S_{Au}}{V_{Au}} \frac{F_{S,Au-\varepsilon}(\gamma_{s,Au,n=1}, \gamma_{s,Ag,n=1})}{F_{V,Au}(\gamma_{s,Au,n=1}, \gamma_{s,Ag,n=1})} + \frac{3}{4} \times v_{F(Ag)} \frac{S_{Ag}}{V_{Ag}} \frac{F_S(\gamma_{s,Au,n=1}, \gamma_{s,Ag,n=1})}{F_V(\gamma_{s,Au,n=1}, \gamma_{s,Ag,n=1})} \right)$$

$$\gamma_{s,Ag,n=2} = \frac{3}{4} \times v_{F(Ag)} \frac{S_{Ag}}{V_{Ag}} \frac{F_{S,Ag,ext}(\gamma_{s,Au,n=1}, \gamma_{s,Ag,n=1})}{F_V(\gamma_{s,Au,n=1}, \gamma_{s,Ag,n=1})} + \frac{|\Delta E_F|}{\bar{E}_F} \frac{1}{2} \left( \frac{3}{4} \times v_{F(Au)} \frac{S_{Au}}{V_{Au}} \frac{F_{S,Au-\varepsilon}(\gamma_{s,Au,n=1}, \gamma_{s,Ag,n=1})}{F_{V,Au}(\gamma_{s,Au,n=1}, \gamma_{s,Ag,n=1})} + \frac{3}{4} \times v_{F(Ag)} \frac{S_{Ag}}{V_{Ag}} \frac{F_{S,Ag+\varepsilon}(\gamma_{s,Au,n=1}, \gamma_{s,Ag,n=1})}{F_{V,Ag}(\gamma_{s,Au,n=1}, \gamma_{s,Ag,n=1})} \right)$$

$$\gamma_{s,Au,n=3} = \frac{|\Delta E_F|}{\bar{E}_F} \frac{1}{2} \left( \frac{3}{4} \times v_{F(Au)} \frac{S_{Au}}{V_{Au}} \frac{F_{S,Au-\varepsilon}(\gamma_{s,Au,n=2}, \gamma_{s,Ag,n=2})}{F_{V,Au}(\gamma_{s,Au,n=2}, \gamma_{s,Ag,n=2})} + \frac{3}{4} \times v_{F(Ag)} \frac{S_{Ag}}{V_{Ag}} \frac{F_S(\gamma_{s,Au,n=2}, \gamma_{s,Ag,n=2})}{F_V(\gamma_{s,Au,n=2}, \gamma_{s,Ag,n=2})} \right)$$

$$\gamma_{s,Ag,n=3} = \frac{3}{4} \times v_{F(Ag)} \frac{S_{Ag}}{V_{Ag}} \frac{F_{S,Ag,ext}(\gamma_{s,Au,n=2}, \gamma_{s,Ag,n=2})}{F_V(\gamma_{s,Au,n=2}, \gamma_{s,Ag,n=2})} + \frac{|\Delta E_F|}{\bar{E}_F} \frac{1}{2} \left( \frac{3}{4} \times v_{F(Au)} \frac{S_{Au}}{V_{Au}} \frac{F_{S,Au-\varepsilon}(\gamma_{s,Au,n=2}, \gamma_{s,Ag,n=2})}{F_{V,Au}(\gamma_{s,Au,n=2}, \gamma_{s,Ag,n=2})} + \frac{3}{4} \times v_{F(Ag)} \frac{S_{Ag}}{V_{Ag}} \frac{F_{S,Ag+\varepsilon}(\gamma_{s,Au,n=2}, \gamma_{s,Ag,n=2})}{F_{V,Ag}(\gamma_{s,Au,n=2}, \gamma_{s,Ag,n=2})} \right)$$

...

$$\gamma_{s,Au,n'} = \frac{|\Delta E_F|}{\bar{E}_F} \frac{1}{2} \left( \frac{3}{4} \times v_{F(Au)} \frac{S_{Au}}{V_{Au}} \frac{F_{S,Au-\varepsilon}(\gamma_{s,Au,n'-1}, \gamma_{s,Ag,n'-1})}{F_{V,Au}(\gamma_{s,Au,n'-1}, \gamma_{s,Ag,n'-1})} + \frac{3}{4} \times v_{F(Ag)} \frac{S_{Ag}}{V_{Ag}} \frac{F_S(\gamma_{s,Au,n'-1}, \gamma_{s,Ag,n'-1})}{F_V(\gamma_{s,Au,n'-1}, \gamma_{s,Ag,n'-1})} \right)$$

$$\gamma_{s,Ag,n'} = \frac{3}{4} \times v_{F(Ag)} \frac{S_{Ag}}{V_{Ag}} \frac{F_{S,Ag,ext}(\gamma_{s,Au,n'-1}, \gamma_{s,Ag,n'-1})}{F_V(\gamma_{s,Au,n'-1}, \gamma_{s,Ag,n'-1})} + \frac{|\Delta E_F|}{\bar{E}_F} \frac{1}{2} \left( \frac{3}{4} \times v_{F(Au)} \frac{S_{Au}}{V_{Au}} \frac{F_{S,Au-\varepsilon}(\gamma_{s,Au,n'-1}, \gamma_{s,Ag,n'-1})}{F_{V,Au}(\gamma_{s,Au,n'-1}, \gamma_{s,Ag,n'-1})} + \frac{3}{4} \times v_{F(Ag)} \frac{S_{Ag}}{V_{Ag}} \frac{F_{S,Ag+\varepsilon}(\gamma_{s,Au,n'-1}, \gamma_{s,Ag,n'-1})}{F_{V,Ag}(\gamma_{s,Au,n'-1}, \gamma_{s,Ag,n'-1})} \right)$$

$n' \to \infty$

Such an iterative procedure was applied by us in this study and did well.

## 1.2 Monometallic NR.

For monometallic NRs, the procedure is simpler – we have only one metal and only one interface. For example, we can set $|\Delta E_F| \to 0$ and obtain, like in Refs. [3,4]:

$$\gamma_{s,\beta} = \frac{3}{4} \times v_{F(\beta)} \frac{S_\beta}{V_\beta} \frac{F_{S,\beta,\text{ext}}(\gamma_{s,\beta,\text{n'-1}}, \gamma_{s,\beta,\text{n'-1}})}{F_V(\gamma_{s,\beta,\text{n'-1}}, \gamma_{s,\beta,\text{n'-1}})}$$

$\beta = $ Au or Ag

To conclude, we give below a few examples of the calculated sets of the quantum dissipation parameters in monometallic and bi-metallic NRs.

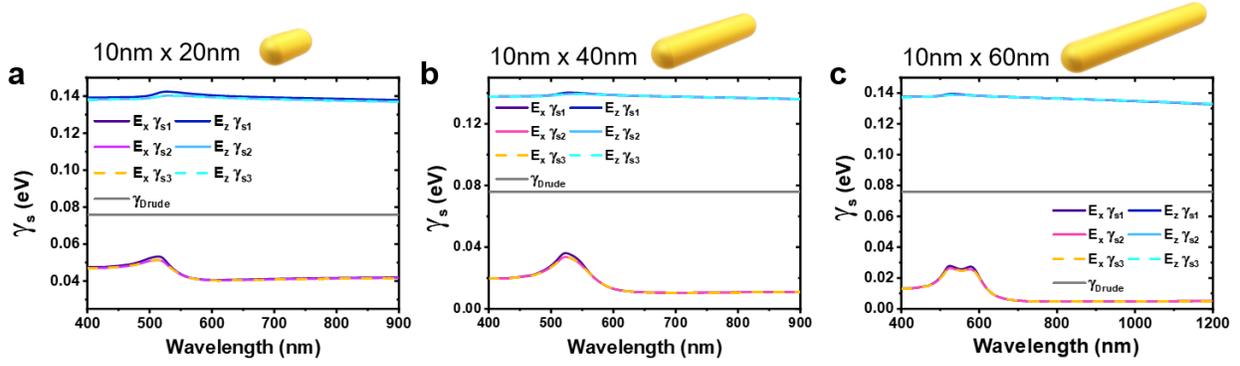

Figure S3. The surface-scattering broadening parameter for gold nanorods of different sizes.

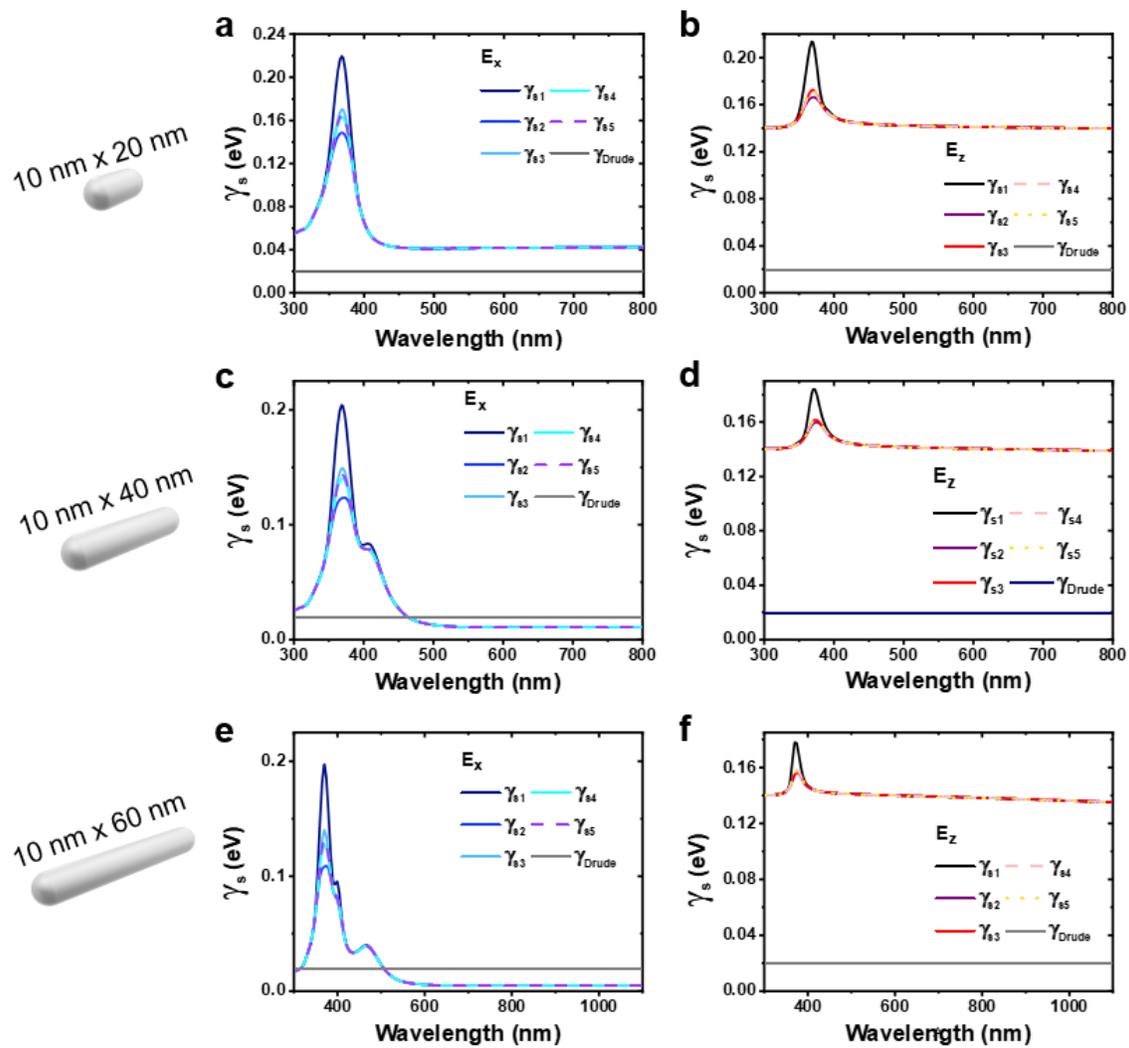

Figure S4. The surface-scattering broadening parameter for silver nanorods of different sizes.

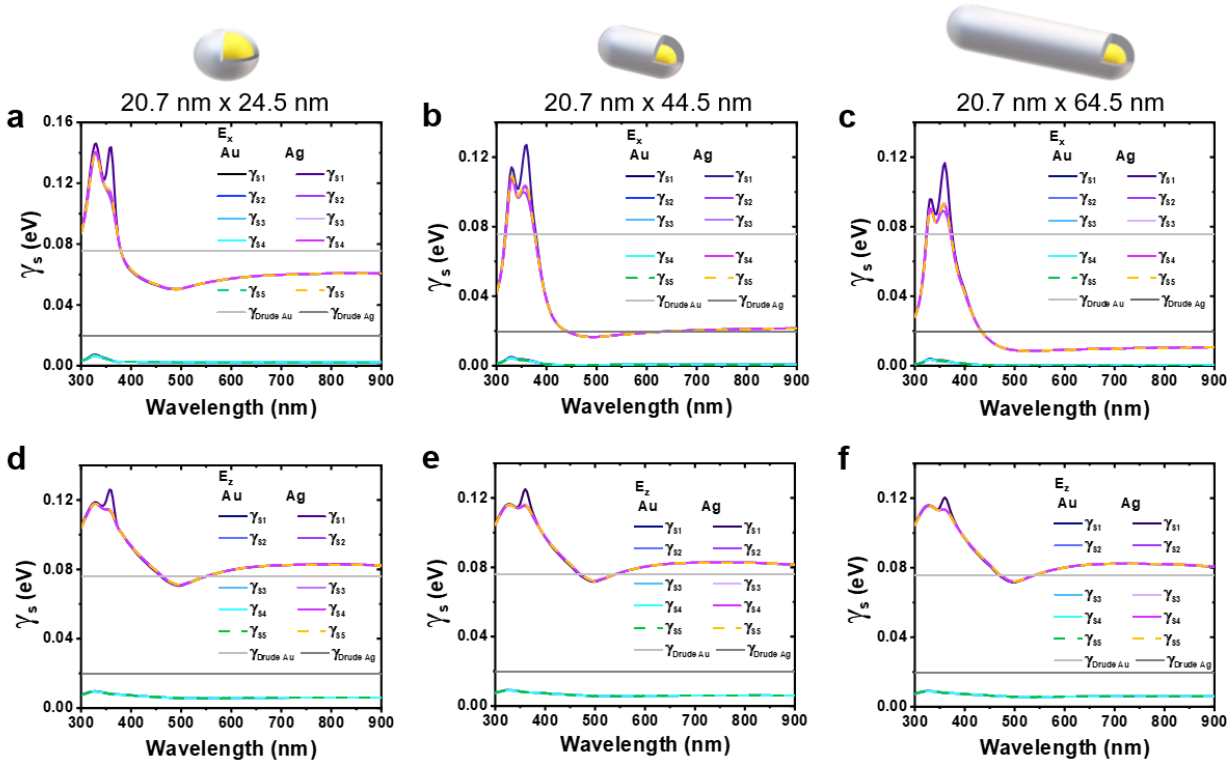

Figure S5. The surface-scattering broadening parameter for core-shell nanorods of different sizes.

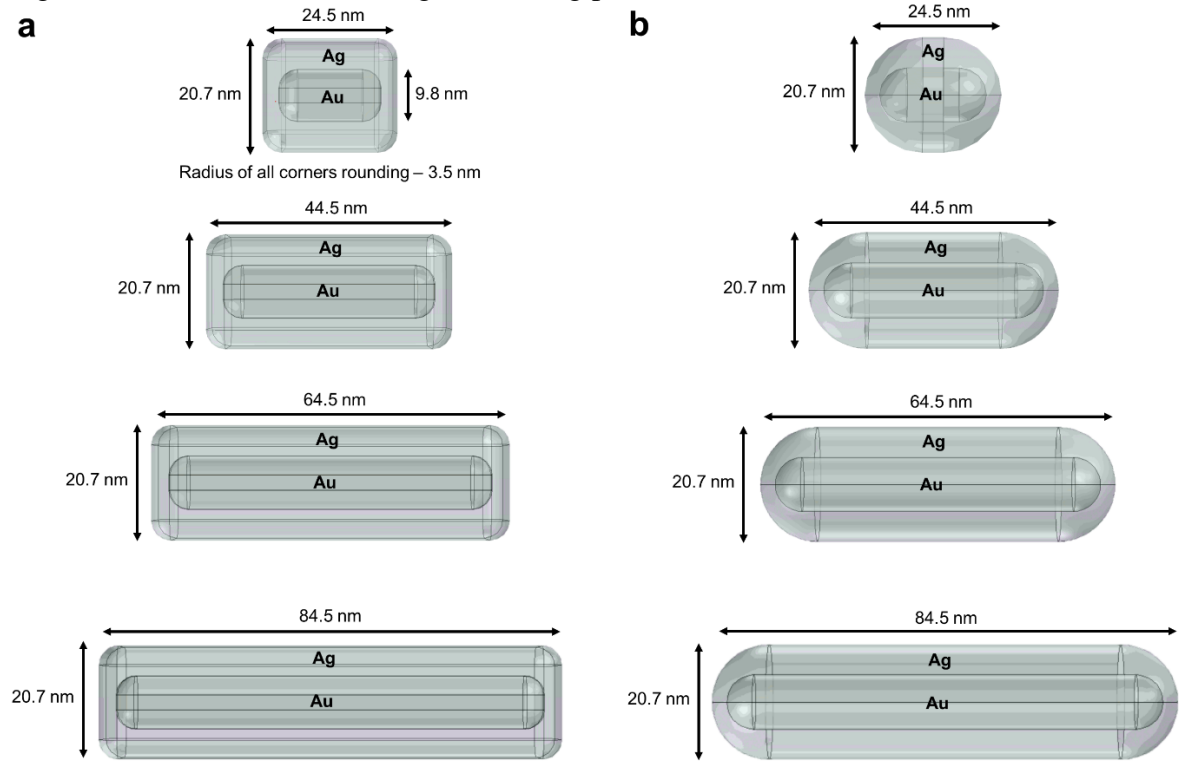

Figure S6. Shapes and calculation parameters for Au@Ag nanocuboids (a) and nanorods (b).

Table S2. Sizes and volumes of the Au@Ag nanocuboids and nanorods

| Size | $V_{Ag}$ (nm$^3$) | $V_{Au}$ (nm$^3$) | Ratio |
|---|---|---|---|
| NR 24.5x20.7 | 4690 | 1189 | 3.94 |
| Cuboid 24.5x20.7 | 8362 | 1486 | 5.63 |
| NR 44.5x20.7 | 12648 | 2708 | 4.67 |
| Cuboid 44.5x20.7 | 15013 | 3192 | 4.7 |
| NR 64.5x20.7 | 19374 | 4215 | 4.59 |
| Cuboid 64.5x20.7 | 26577 | 4911 | 5.41 |
| NR 84.5x20.7 | 26098 | 5721 | 4.56 |
| Cuboid 84.5x20.7 | 34934 | 6620 | 5.27 |

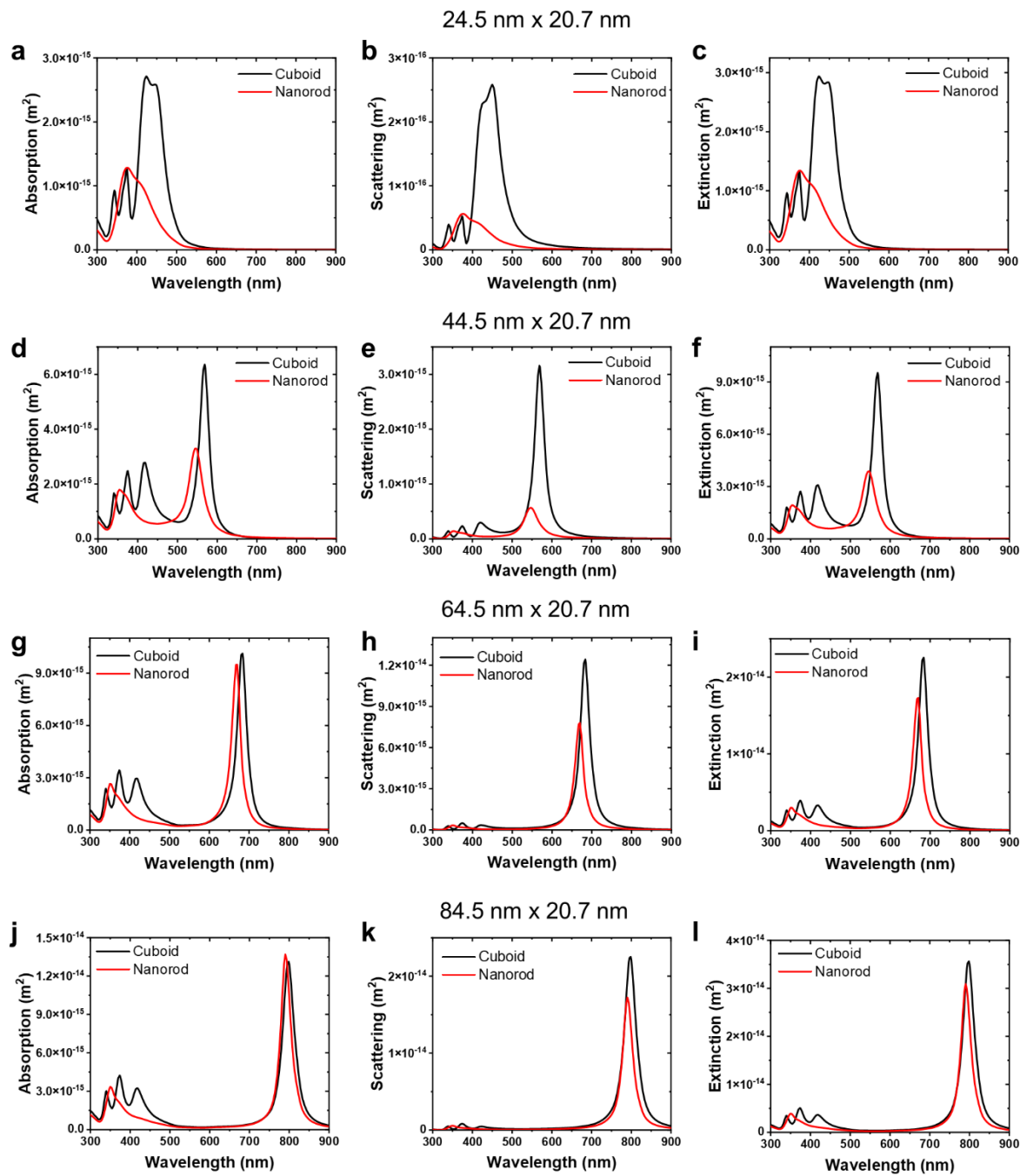

Figure S7. Absorption (a,d,g,j), scattering (b,e,h,k) and extinction (c,f,i,l) of the Au@Ag core-shell nanocuboids and nanorods of different sizes.

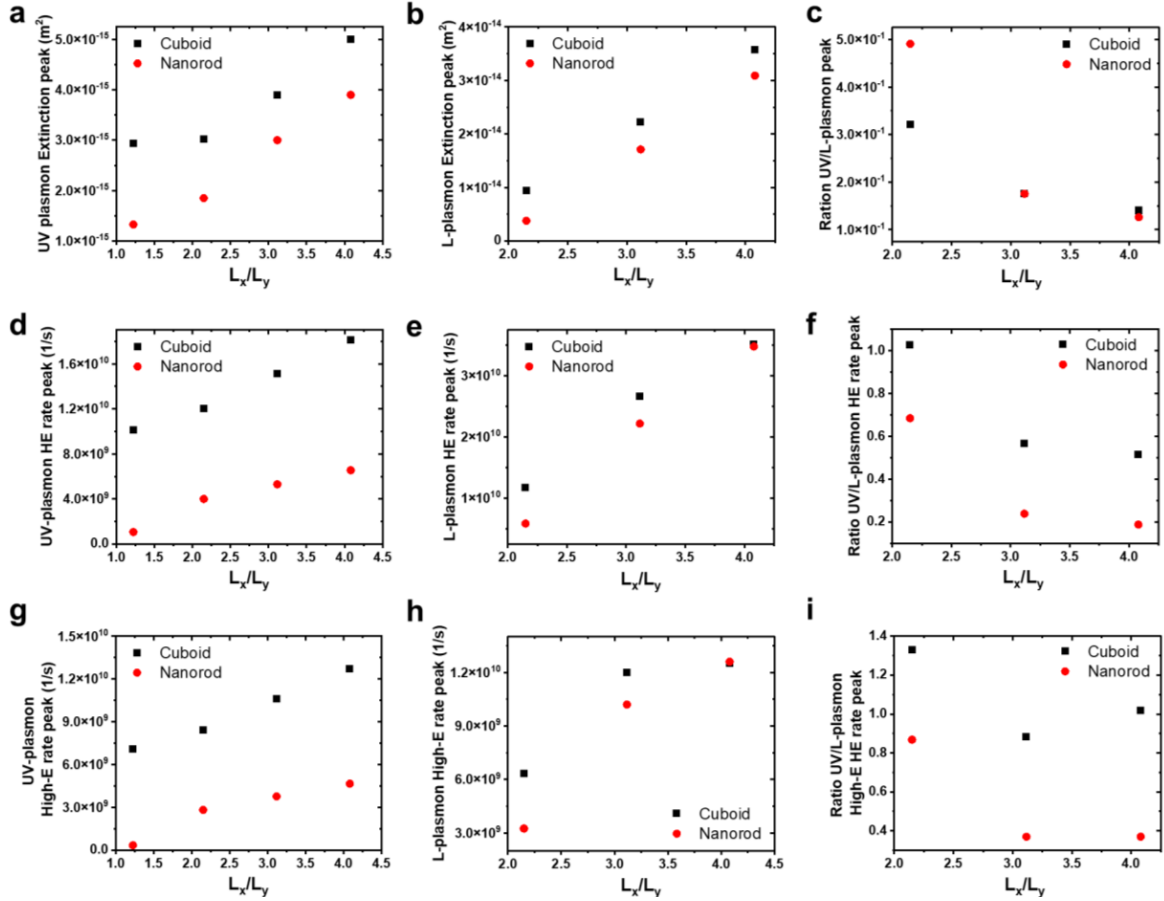

Figure S8. Energetic and optical properties of the Au@Ag core-shell nanocuboids and nanorods depending on aspect ratio. Dependencies of the extinction (a-c), HE generation rates (d-f) and High-E HE generation rates peaks intensities of the UV (a,d,g) and L-plasmon (b,e,h) and their ratio (c,f,i) on aspect ratio.

## S2. Efficiencies: internal vs. external

Here, we present few examples of **the experimental external efficiencies**:
- gold nanoparticles in nanoporous $TiO_2$ film, photo-catalysis: $Eff_{transfer}$ (incident photon-to-conversion efficiency) ~ 1 %.[5]

- Au-tipped CdS nanorods, with AuNPs size from 1.6 nm to 5.5 nm: $Eff_{HE,external}$ ($Eff_{transfer}$ from Au to a semiconductor, CdSe nanorods) =1-18 %.[6]

- Ag NP sensitized nanoporous anatase $TiO_2$ thin films. AgNPs, with AgNPs size from 1.7 nm to 5.9 nm: $Eff_{external}$ ($Eff_{transfer}$ to a semiconductor)=28-53 %.[7]

Our **computed internal** efficiency of intraband HEs at the plasmon peak for:

AuNP~ 20% (Fig. 4d)

AuNR~ 45% (Fig. 4h)

AgNR ~ 50% (Fig. 5f)

Au@AgNR: the efficiencies between the Au and Ag cases.

The internal efficiency of d-holes should be high since it is due to the interband transitions, as we see directly from the empirical dielectric function. Efficiency of d-holes ~ 0.6 for both Au and Ag at the T-plasmon peak (Fig. 10a,b) and is so for AuNPs.[3]

Regarding the issue of how to evaluate the external efficiency of HEs (i.e., $Eff_{HE,external}$), this problem is complex.[1,3] However, to treat it, one can use a simplified rate model,[3] which is shown in Figure S9 below. In this model, the transfer efficiency (or the transfer probability) for the HEs reads:

$$Eff_{transfer} = \frac{\gamma_{trans}}{\gamma_{trans} + \gamma_{rel}}$$

where $\gamma_{trans} = 1/\tau_{trans}$ is the rate of charge transfer from the nanoparticle to the environment, and $\gamma_{rel} = \bar{\gamma}_{ph} + \bar{\gamma}_{ee} = 1/\tau_{rel}$ is the electron's relaxation rate; $\bar{\gamma}_{ph}$ is the average phonon-induced relaxation rate, and $\bar{\gamma}_{ee}$ is relaxation rate due to the electron-electron collisions.

Then, we can define external efficiency of over-barrier generation as

$$Eff_{HE,external} = Eff_{transfer} \cdot Eff_{high\text{-}E\ HE}$$

where $Eff_{high\text{-}E\ HE}$ was defined in details in main text. We also illustrate the model for $Eff_{HE,external}$:

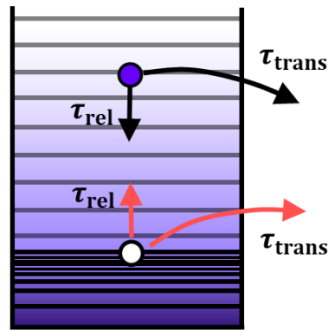

Figure S9. Rate model illustrating fast electron−electron relaxation of HEs and charge transfer from a nanoparticle.

## S3. Short review of experiments on Au@Ag NRs

**S3.1 Synthesis and plasmon resonances:** From the first reports of the synthesis,[8–11] AuNRs attract much attention due to their optical properties caused by the presence of two plasmon resonances (transversal and longitudinal) and strong electromagnetic fields concentrated near the AuNRs tips. The position of the longitudinal LSPR varies linearly with the aspect ratio of GNRs, making the latter a tunable and versatile platform for applications in biomedical technologies, plasmon-enhanced spectroscopy, and optoelectronic devices.[10,12,13] Au NRs are usually synthesized by a seed-mediated growth method and stabilized by CTAB (cetyltrimethylammonium bromide) molecules that form on their surface a dense insulating bilayer with the thickness of 3.2 nm.[14,15] Increasing the AR of GNR during the synthesis allow tuning the longitudinal localized surface plasmon resonance (LSPR) from ~620 nm to the near-IR range, while the growth of immediate nanorod dimensions leads to the rise of scattering contribution in the extinction spectrum. Moreover, colloidal synthesis and further chemical treatment provide the access to their surface and electronic properties.

The hot-electron generation in AuNRs was studied theoretically[4] and experimentally[16–19] in the context of photocatalytic and photoluminescent applications. Especially, AuNRs have longitudinal LSPR dominated by absorption (Figure 3) and well separated from the interband region, allowing to estimate separately contributions of intraband and interband transitions to hot-carriers generation.[17] However, the reported experimental efficiencies (~2% in [16]) were not high, which is smaller than we observe in theoretical predictions (Figure 4), and we discuss it in previous section S2.

An important point is that colloidal Au NRs with have high monodispersity and stability are widely available nowadays, while reproducible and high yield synthesis of AgNRs is still challenging due to their fast oxidation and sulfidation.[20–22] Also, the size and shape of the synthesized nanoparticles are harder to control compared to gold nanoparticles, and they need additional protective layers.[22–24] However, there were few successful reports on the Ag nanorods and nanobars photochemical[25,26] or polyol synthesis.[27,28] Although, the plasmonic properties of AgNRs are of greater interest than of AuNRs, because of stronger local field enhancement and coverage of broader spectral region, reports on their use are limited to SERS, plasmonic sensors, and corresponding bioapplications.[24,26] Also, in some articles synthesis of AgNRs is directly described with gold nanorods as seeds to grow silver shells since it provides higher stability and tunability of the nanoparticles.[5,6] Higher stability of the Ag covered Au nanoparticles are usually explained by few reasons. The large electronegativity of gold can reduce the electron density on the Ag part and therefore make the Ag layer difficult to oxidize. Also, interaction of Ag atoms with Au can induce charge redistribution from the 4s orbitals to the 3d orbitals thus increasing the resistance of Ag to oxidation.[29] Thus, hybrid core-shell structures, such as Au@Ag NRs, are already under active investigation not only because of combining the properties of the two materials but also because of new effects and excited modes. Next, we outline few major motivations of this study of Au@Ag NRs:

(1) While purely Ag NRs seem to be less stable or need spatial coating[1], higher stability of Au@Ag NRs is well documented (like in ref. [29–31]), when Ag shell is overgrown on the highly stable Au-NR core.

(2) Easy to fabricate and control since the starting point is Au NRs, which synthesis is well developed and widely available.

(3) Au@Ag NRs provide wide tunability of the relative composition,[32] sizes and the plasmon wavelength and also have higher variety of the plasmonic modes covering larger spectral range,[33,34] while AgNRs and AuNRs provides mainly two plasmonic modes.

(4) The presence of the Ag-slab plasmon in the UV (350 nm), which is not present in the Ag NRs.

(5) Au@Ag NRs have strong and blue-shifted plasmon resonances compared to AuNRs. Although, Ag NRs should still have stronger resonances as compared to Au@Ag NRs, the core-shell structure possesses unique properties described above.

(6) Previous successful reports on the use of Au@Ag NRs in multiple applications.[29–31,35]

## S3.2 Applications

The geometric models developed by us were used in Refs. [30,31] (Figure S10 and S11). Our electromagnetic calculations described well those data. However, our calculations were done only for a few limited sets of geometrical parameters and also for the interaction n=0 (i.e., we did not apply the full 2-$\gamma_s$ formalism). There were number of unsolved problems in Refs. [30,31]:

(1) Study of Ref. [31] revealed the ultra-strong generation of HEs in the UV at about 350 nm (Figure S10c), and an explanation for this observation has been missing so far. This paper explains it well, involving the slab-like Ag-based plasmon resonance.

(2) The origin and the appearance of the UV Ag plasmon (slab-like plasmon) at 350 nm.

(3) A new formalism with 2-$\gamma_s$ for a bi-metal nanostructure was needed in those papers, but it was not developed yet. It is an important development for the description of bi-metallic (or tri-metallic, or even more complex) nanostructures.

(4) The above papers did not look at the retardation/electromagnetic effects on HEs that start playing a major role for long NRs or for NRs with a thick Ag-shell.

The current paper by us has solved the above problems.

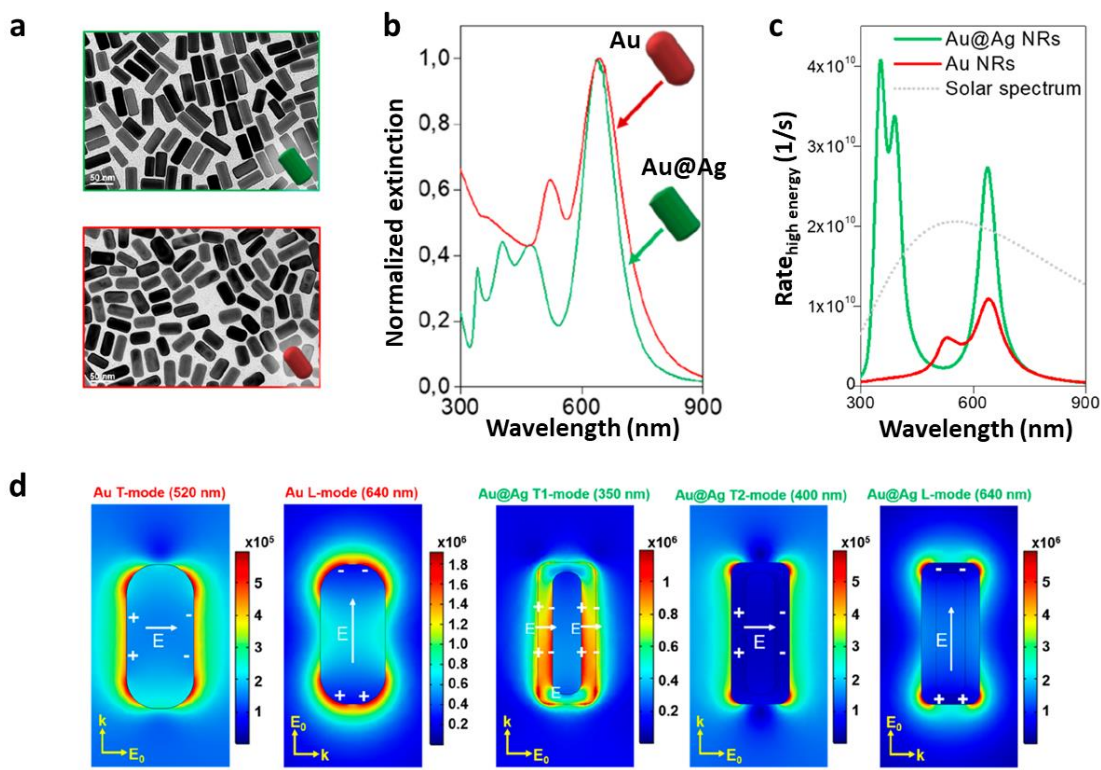

Figure S10. TEM images (a) and experimental extinction spectra (b) of the Au NRs (red) and the Au@Ag NRs (green), aspect ratios – 2.1 and 2.3, respectively, (c) Calculated rates of generation of hot electrons for the Au and Au@Ag NRs. (d) Near-field enhancement maps of the plasmonic modes of the Au and Au@Ag NRs. Adapted with permission from [31]. Copyright 2020 American Chemical Society.

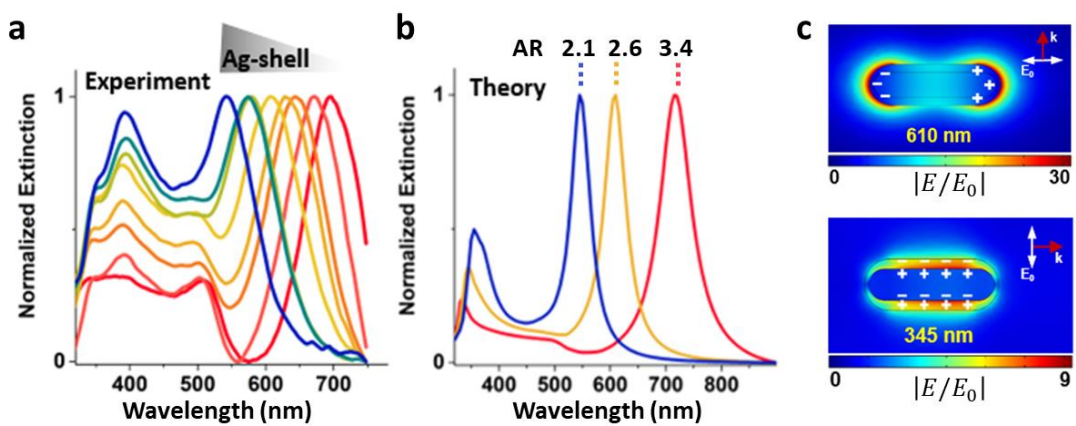

Figure S11. (a) Experimental extinction spectra of Au@Ag NRs@DNA synthesized using different $AgNO_3$ concentrations (10−40 mM) resulting in Ag shell thicknesses from ∼1.1 (red line) to ∼5.4 nm (blue line). (b) Calculated theoretical extinction spectra of Au@Ag NRs in water for the Ag shells of 1.1 nm (red line), 3.1 nm (yellow line), and 5.4 nm (blue line). (c) Near-field

enhancement maps of plasmonic modes of Au@Ag NRs. Adapted with permission from [30]. Copyright 2020 American Chemical Society.